\documentclass[fleqn,usenatbib,useAMS]{mnras}
\usepackage[T1]{fontenc}
\usepackage[title]{appendix}
\usepackage{pslatex}
\usepackage{float}
\usepackage{ae,aecompl}
\usepackage{gensymb}
\usepackage[section]{placeins}
\usepackage{amsmath}
\usepackage{relsize}
\usepackage{graphicx}	% Including figure files
\usepackage{amssymb}	% Extra maths symbols
\usepackage{multicol}        % Multi-column entries in tables
\usepackage{bm}		% Bold maths symbols, including upright Greek
\usepackage{pdflscape}	% Landscape pages

\usepackage{newtxtext,newtxmath}

\title[Analysis of mCP stars]{Analysis of eight magnetic chemically peculiar stars with rotational modulation}
\author[O. Kobzar et al.] {
        {\parbox{\linewidth}{\flushleft O. Kobzar$^1$, V. Khalack$^1$, D. Bohlender$^2$, G. Mathys$^3$, M. E. Shultz$^4$, D. M. Bowman$^5$, E. Paunzen$^6$, C. Lovekin$^7$, A. David-Uraz$^{8,9}$,  J. Sikora$^{10}$,  P. Lampens$^{11}$, O. Richard$^{12}$}}\\
        \\
          $^1$D\'epartement de Physique et d'Astronomie, Universit\'e de Moncton, Moncton, N.B., Canada E1A 3E9\\
          $^{2}$National Research Council of Canada, Herzberg Institute of Astronomy and Astrophysics, 5071 West Saanich Road, Victoria, BC, Canada V9E 2E7\\
          $^3$European Southern Observatory, Alonso de Cordova 3107, Vitacura, Santiago, Chile \\
          $^4$Department of Physics and Astronomy, University of Delaware, 217 Sharp Lab, Newark, DE 19716, USA\\
          $^5$Institute of Astronomy, KU Leuven, Celestijnenlaan 200D, B-3001 Leuven, Belgium\\
          $^6$Department of Theoretical Physics and Astrophysics, Masaryk University, Kotl\'a\v{r}sk\'a 2, 611\,37 Brno, Czech Republic\\
          $^7$Department of Physics, Mount Allison University, Sackville, N.B., Canada E4L 1E6\\
          $^{8}$Department of Physics and Astronomy, Howard University, Washington, DC 20059, USA \\ 
          $^{9}$Center for Research and Exploration in Space Science and Technology, and X-ray Astrophysics Laboratory, NASA/GSFC, Greenbelt, MD 20771, USA \\
          $^{10}$Department of Physics \& Astronomy, Bishop's University, Sherbrooke, QC J1M 1Z7, Canada\\
          $^{11}$Royal Observatory of Belgium, Ringlaan 3, B-1180 Brussels, Belgium\\
          $^{12}$Laboratoire Univers et Particules de Montpellier, Universit\'{e} de Montpellier, B\^{a}t 13- CC072, Place Eug\`{e}ne Bataillon, 34095 Montpellier, France
            }
\date{Accepted ???.
      Received ???;
      in original form ???}
\pubyear{2022}

\begin{document}
\label{firstpage}
\pagerange{\pageref{firstpage}--\pageref{lastpage}} 
\maketitle

\begin{abstract}

Since the end of 2018, the Transiting Exoplanet Survey Satellite (\textit{TESS}) has provided stellar photometry to the astronomical community. We have used \textit{TESS} data to study rotational modulation in the light curves of a sample of chemically peculiar stars with measured large-scale magnetic fields (mCP stars). In general, mCP stars show inhomogeneous distributions of elements in their atmospheres that lead to spectroscopic (line profile) and photometric (light curve) variations commensurate with the rotational period. We analyzed the available \textit{TESS} data from 50 sectors for eight targets after post-processing them in order to minimize systematic instrumental trends. Analysis of the light curves allowed us to determine rotational periods for all eight of our targets. For each star, we provide a phase diagram calculated using the derived period from the light curves and from the available measurements of the disk-averaged longitudinal magnetic field $\langle B_{\rm z}\rangle$. In most cases, the phased light curve and $\langle B_{\rm z}\rangle$ measurements show consistent variability.
Using our rotation periods, and global stellar parameters derived from fitting Balmer line profiles, and from Geneva and Str{\"o}mgren-Crawford photometry, we determined the equatorial rotational velocities and calculated the respective critical rotational fractions $v_{\rm eq}/v_{\rm crit}$. We have shown from our sample that the critical rotational fraction decreases with stellar age, at a rate consistent with the magnetic braking observed in the larger population of mCP stars.
\end{abstract}


%\begingroup
%\let\clearpage\relax
%\tableofcontents
%\endgroup
%\newpage

\section{Introduction}

\begin{table*}
\begin{center}
\caption{Stellar parameters collected from the literature and derived from our analysis of observed data for the ApBp stars studied in this work. Column~(1) presents the HD number, %with alternative names in the title for each star in the Sector\ref{results}, 
column~(2) shows the spectral type of each star inferred from recent studies. %taken from the most recent study, 
Columns~(3), (4), (5), and (6) provide data for effective temperature and surface gravity collected from the \textit{TESS} Input Catalog (TIC)$^1$ and derived from fitting the Balmer line profiles respectively. Column~(7) and (8) present rotational periods %of rotation obtained from the LC analysis, 
extracted from the literature and derived from the analysis of light curves. Columns~(9) and (10), and (11) and (12) show v$\sin{i}$ and radial velocity inferred from the literature and derived from fitting the Balmer line profiles, respectively. } 
%with two subcategories of the values defined  survey and Balmer line profiles fitting procedure.}
\label{tab1}
\def\arraystretch{0.98}
\setlength{\tabcolsep}{3pt}
\begin{tabular}{lccccccccccc}
\hline
Name& Spectral type&\multicolumn{2}{c}{$T_{\rm eff}$ (K)} &\multicolumn{2}{c}{
$\log{g}$}&\multicolumn{2}{c}{Period (d)}&\multicolumn{2}{c}{v$\sin{i}$ (km~s$^{-1}$)}
&\multicolumn{2}{c}{v$_{\rm r}$  (km~s$^{-1}$)}\\

&SIMBAD &TIC$^1$ & This study &TIC$^1$ & This study & Published & This study & SIMBAD & This study& SIMBAD &This study \\
& & & Balmer & & Balmer & & & & Balmer & & Balmer\\
\hline
HD10840 & B9$^{2}$ & 11663$\pm$317& - & 4.11$\pm$0.07& -&  2.097679(7)$^{6}$ & 2.0976858(2)  &35.0$\pm$5.0$^{14}$& - & 19.4$\pm$2.1$^{21}$& - \\
HD22920 &B8II$^{3}$ &13800$\pm$200&13678$\pm$200& - &3.77$\pm$0.20& 3.9472(1)$^{7}$ & 3.947225(2)  &37.0$\pm$5.0$^{15}$ &33.0$\pm$5.0 &18.0$\pm$4.0$^{15}$&18.0$\pm$3.0\\
HD24712 &A9Vp$^{4}$ &7242$\pm$129& 7290$\pm$200& 4.18$\pm$0.08 &4.06$\pm$0.20&  12.461(1)$^{8}$ & 12.45862(5)  &6.6$\pm$0.6$^{16} $&9.0$\pm$3.0&23.2$\pm$0.4$^{22}$&22.0$\pm$2.0\\
HD38170 &B9$^{5}$ & 10000$\pm$280 & 9470$\pm$200 & 3.80 $\pm$ 0.08 &3.65$\pm$0.20& 2.76618(4)$^{9}$ & 2.766116(2) & 65.0$\pm$9.0$^{17}$ & 65.0$\pm$3.0 & 36.3$\pm$0.6$^{22}$& 33.0$\pm$2.0 \\
HD63401 & B9$^{5}$ & - & 13356$\pm$200 & - & 4.06$\pm$0.20& 2.41(2)$^{10}$ & 2.414474(1) & 52.0$\pm$4.0$^{18}$ & 52.0$\pm$4.0 & 22.0$\pm$1.4$^{22}$ & 26.0$\pm$4.0\\
HD74521 & A1Vp$^{6}$ & 12188$\pm$146 & 10600$\pm$200 & - & 3.47$\pm$0.20&7.0501(2)$^{11}$ & 7.05010(5)  & 19.0$\pm$4.6$^{19}$& 18.0$\pm$3.0 & 27.5$\pm$1.4$^{22}$ & 24.0$\pm$3.0 \\
HD77314 & A2$^{5}$ & 9253$\pm$372& 11437$\pm$200 &3.69$\pm$0.10& 3.99$\pm$0.20& 2.86445(8)$^{12}$ & 2.864325(1)  &-& 47.0$\pm$3.0&-&  -4.0$\pm$1.0 \\
HD86592 & A0$^{5}$ & 8129 $\pm$145& 7804$\pm$200 &4.18$\pm$0.07& 3.83$\pm$0.20&2.8867$^{13}$ & 2.88657(3)  & 16.2$\pm$2.0$^{20}$ & 27.0$\pm$5.0 &12.7$\pm$0.3$^{20}$& 13.0$\pm$1.0\\
\hline
\end{tabular}
\end{center}
Note: {{$^1$}\citet{Stassun2018, Stassun2019}, {$^{2}$}\citet{Renson1992}, {$^{3}$}\citet{Houk1999}, {$^{4}$}\citet{Abt1995}, {$^{5}$}\citet{Cannon1993}, {$^{6}$}\citet{Sikora2019}, {$^{7}$}\citet{Shultz2022}, {$^{8}$}\citet{Bagnulo1995}, {$^{9}$}\citet{David-Uraz2021}, {$^{10}$}\citet{Hensberge1976}, {$^{11}$}\citet{Dukes2018}, {$^{12}$}\citet{Bernhard2020}, {$^{13}$}\citet{Babel1997}}, {$^{14}$}\citet{Bailey2013},  {$^{15}$}\citet{Khalack2015}, {$^{16}$}\citet{Sikora2019_2}, {$^{17}$}\citet{Royer2002}, {$^{18}$}\citet{Bailey2014},
{$^{19}$}\citet{Mathys1995}, {$^{20}$}\citet{Babel1997}, {$^{21}$}\citet{Levato1996}, {$^{22}$}\citet{Gontcharov2006}
\end{table*}

Chemically peculiar (CP) stars on the upper main sequence (MS) are identified through abnormally strong absorption lines of some chemical elements. In particular ApBp stars, which were classified as CP2 by \citet{Preston1974}, are characterized by strong lines of metals such as Si, Cr, Fe, Sr, % Eu, 
and rare earth elements. Such peculiarities that occur in stellar atmospheres might be the result of the competition between gravitational settling and radiative acceleration \citep{Michaud1970}, leading to a relative movement between the different ions of the plasma known as atomic diffusion 
\citep{Alecian2010}. \par
\citet{Preston1974} mentioned that the group of magnetic ApBp stars shows relatively slow rotation compared to normal stars of A and B spectral types with the same effective temperatures.
%range which possess a significant magnetic field \citep{Bychkov2003, Buysschaert2018, Shultz2022}. 
Magnetic fields detected in ApBp stars \citep[e.g.][]{Bychkov2003, Buysschaert2018, Shultz2022} are thought to have a fossil origin \citep{Neiner2015} and are known to range in strength from a few hundred G up to a few tens of kG \citep{Auriere2007, Shultz2019, Sikora2019b}. The influence of magnetic fields in ApBp stars is quite significant. It prevents rotational mixing and therefore amplifies atomic diffusion \citep{Alecian2010} producing an inhomogeneous redistribution of elements in stellar atmospheres.  It is thought that the magnetic field stabilizes the stellar atmosphere, enabling diffusion to play a greater role as compared, for example, to Am stars \citep{Michaud2015}, which are generally not known to host strong, organized fields at their surface. The magnetic fields of most ApBp stars tend to be predominantly dipolar \citep{Kochukhov2019}.  
%The presence of chemical spots can be detected %in a way that is visible 
%through the analysis of spectrophotometric \citep{Krtika2015}, photometric and magnetic field data  that often show periodic variations in the case of ApBp stars. 
%More complex magnetic field geometries were investigated for magnetic CP stars in Magnetic Doppler Imaging (MDI) studies \citep{Kochukhov2002, Pishkunov2002} which explain the interrelation between magnetic field and chemical spots for these stars \citep{Kochukhov2011, Silvester2017}. \par
%Change the next phrase!
%Since

In general, the rotational and magnetic dipole axes are not aligned in CP stars.% the abundance spots of peculiar metals
%gathered around the magnetic poles 
To first order, the distribution of the abundance spots over the stellar surface is governed by the magnetic geometry (\citealt{Alecian2019, Alecian2021}). These spots result in flux redistribution which is translated to the appearance of brightness spots leading to the modulation of light curves by stellar rotation \citep{Krtika2015}. 
%The location of peculiar abundance spots can be described by employing the oblique magnetic rotator (OMR) model introduced by \citet{Stibbs1950}. %, the oblique magnetic rotator (OMR) model was investigated in the 1960~s, and after repeated improvements and observations is widely accepted to describe the magnetic field in stars. 
The description of abundance anomalies in ApBp stars is not an easy task considering that an element's abundance shows both horizontal segregation and vertical stratification in their atmospheres \citep{Leblanc2015, Khalack2017, Khalack2018, Ndiaye2018, Khalack2020}. 
% are much more difficult to describe for this group in the case of non-standard concentration of elements not only in the vertical but also in the horizontal region. Nevertheless, 
The group of mCP stars exhibits variability of spectral line profiles \citep{Krtika2015}, surface brightness, and magnetic field measurements taken at different epochs, that all appear to be modulated by the stellar rotation period \citep{Samus2017}. %Fainter 
Most CP stars have
%remain beside the loop and 
not been identified yet as magnetic since there are no high-quality polarimetric data with high enough signal-to-noise ratio (see for details \citealt{Donati1997, Kochukhov2010, Kochukhov2018})  from which one can detect a significant magnetic field. %A complex magnetic field geometry can complicate even more the detection of magnetic field in the faint CP stars.
%, or axis of the magnetic dipole is directed to Earth,
%or extremely long period of stellar rotation is too long to acquire decent observations \citep{Cunha2019}.\par

%Knowledge of precise value of rotational period and magnetic field structure is required for 
Our high-level project is aimed to build a sample of stars for which large, high-quality, high-resolution spectropolarimetric time series densely sampling the rotational period are possible to obtain. 
These requirements are applicable mainly for fairly bright stars considering that high-resolution spectropolarimetric observation are usually in high demand. We determined to use this spectropolarimetric data for detailed Magnetic Doppler Imaging studies \citep{Kochukhov2002, Pishkunov2002} %hence which lends itself
to investigate the impact of magnetic field strength and geometry on the horizontal and vertical abundance differentiation of chemical elements in the atmospheres of ApBp stars (see e.g. \citealt{Alecian2010, Kochukhov2011, Rusomarov2015, Silvester2017, Alecian2019, Alecian2021}).% can be performed at the foreseen future. %in the following studies. 
We looked for relatively slowly rotating CP stars (P$_{\textrm {rot}}$> 2 days and v$\sin{i}$< 65 km~s$^{-1}$) which may possess a hydrodynamically stable atmosphere. We searched for bright stars ($V <8.0$~mag) with available high-resolution and high signal-to-noise spectra, which are required for the investigation of abundance stratification. %The future studies similar to the aforementioned investigations will %lay a groundwork in order to find answer to the open questions of modern stellar astrophysics: 
%help to understand why abundance patterns differ between the elements, and why the magnetic field geometry influences structure as expected.
A sample of eight objects was selected for this study based on the aforementioned criteria through the preliminary analysis of ApBp stars observed by \textit{TESS} during the first cycle of its mission, which is referred to as the first year of observation \citep{Kobzar2020}. %(see Subsection~\ref{photoTESS} for the details on selection procedure).
The main goal of our study is
%, we aimed 
to confirm the rotation periods of these eight ApBp stars based on the photometric observations provided by the \textit{TESS} mission, and on the available measurements of the mean longitudinal magnetic field determined in this study and extracted from literature (see Table~\ref{appendix}). %polarimetric available spectra, including databases and own observations. Primarily, 

% were selected from the first year of observation of telescope, besides for some objects become possible to increase the timeline by adding data from the third year of observations. Since 
Considering that the magnetic field structure in ApBp stars is stable over at least decades \citep{Silvester2017, Shultz2018}, the measurements of the mean longitudinal magnetic field collected for each selected target should vary on the same timescale as the light curve, which allows one to derive or to confirm the period of stellar rotation with high accuracy.
%In order to check the consistency of distribution of rotation and age with the one observed with the larger population, %thereby serving as a consistency check on the fundamental parameters inferred from spectroscopic modelling.
 %In order to refine %rotational velocity -
%the age dependency of rotation on the MS, %we aimed 
In order to check the consistency of distribution of rotation and age with the one observed with the larger population of ApBp stars,
our second goal is 
to determine global stellar parameters from Geneva and Str{\"o}mgren-Crawford photometry, atmospheric parameters via spectroscopic measurement, and fundamental parameters via available evolutionary models. Results of this study provide the empirical foundation for a future analysis of abundance stratification which will help to understand %better
how the redistribution of chemical elements is related to the stellar 
%with which to look for trends e.g. 
age, rotation, and magnetic field strength, etc.\par
%The process extraction of original data, such as selection of targets from the TESS observation and magnetic field measurements, and data analysis 
Observations and data reduction are described in Section \ref{obs}. Results obtained for individual stars from our sample are presented in Section \ref{results}. Discussion of the derived results and conclusions follow in Section \ref{discus}, and summary is outlined in Section~\ref{summary}.

\section{Observations, data reduction and methods}
\label{obs}

\subsection{\textit{TESS} data}
\label{photoTESS}

\begin{figure}
\centering
\includegraphics[width=\linewidth]{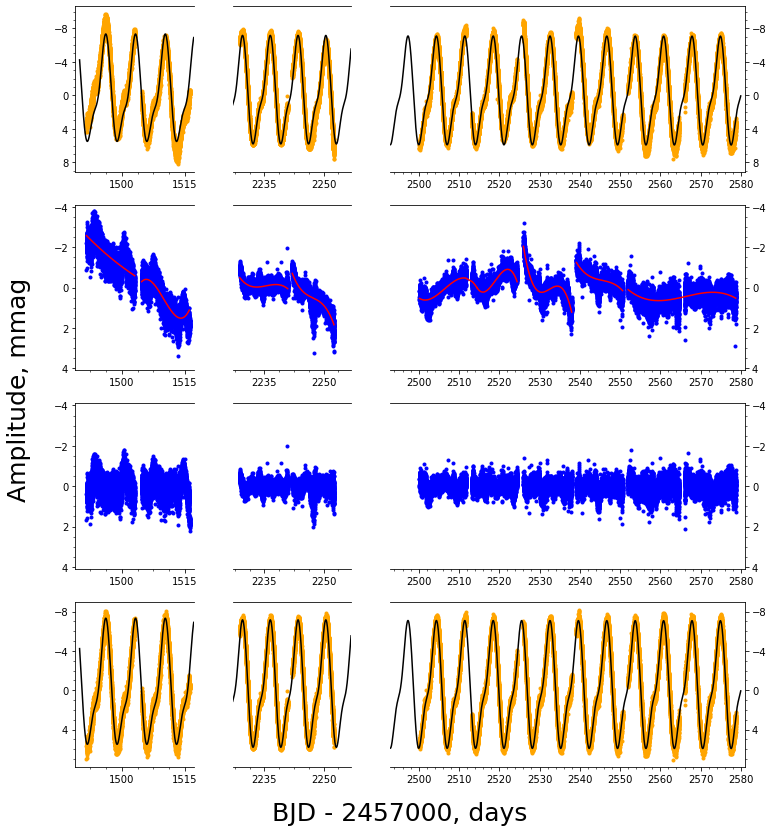}
\caption{Outline of the detrending procedure of the \textit{TESS} light curve (yellow) for the star HD~74521. The first panel from the top presents the original light curve where the green solid line describes harmonic model, the second panel shows the light curve pre-whitened with the inferred rotational frequency and its harmonic(s) where the red line the polynomial fit to the residuals, which is subtracted out to remove instrumental distortion, the third panel contains remaining noise with significantly reduced instrumental distortion, the fourth panel presents the final detrended light curve with green line harmonic model, removing the same polynomial fit as in the previous panel from the original light curve.}
\label{detrending}
\end{figure}
The Transiting Exoplanet Survey Satellite (\textit{TESS}) space mission was launched on 2018 April 18 in order to detect exoplanets via high-precision photometry \citep{Ricker2015}. The data produced by this mission are extremely useful, as they allow the study of not only exoplanets, but variable stars as well. Here we are interested in CP stars that show rotational modulation  and, in some cases (HD~24712), stellar pulsations \citep[roAp type;][]{Kurtz1978} in the presence of a magnetic field \citep{Cunha2019, Holdsworth2021a, Holdsworth2021}. The observing plan for the primary \textit{TESS} mission was to divide the celestial sphere into 26 sectors (with 24$^{\circ}\times$96$^{\circ}$ sky area covered by each sector), so that 13 sectors fall on each ecliptic hemisphere, and overlap near the ecliptic poles. The orbital period of the telescope around the Earth is 13.7~d \citep{Gangestad2013}, which corresponds exactly to half the length of the observation of each sector. One year of observation was dedicated to survey each hemisphere. The photometer works in two observing modes: full frame images (FFIs) from CCDs of each camera, whilst target pixel files (TPFs) collect data only from pixels encompassing pre-selected target stars. \textit{TESS} accumulates data with a 30~min (FFIs) cadence for all stars, %(i.e. taking the full picture),
and with a 2~min (TPFs) cadence for approximately 200,000 objects that were accumulated from the accepted proposals for the primary mission observations \citep{Stassun2018, Stassun2019}.  After Cycle 2 the available modes have been expanded to 20~sec (TPFs) cadence and 10~min (FFIs) cadence observations.\par

%Eight ApBp stars with a long periods of rotation that supposedly have a magnetic field are presented in Table~\ref{tab1}. 
We have used \textit{TESS} data produced by %from  %of the first nine sectors which is in free access 
the Science Processing Operations Center (SPOC)\footnote{https://archive.stsci.edu/hlsp/tess-spoc} pipeline \citep{Jenkins2016, Caldwell2020} and \textit{TESS} light curves\footnote{https://archive.stsci.edu/tess/bulk\_downloads.html} which are available from the Mikulski Archive for Space Telescopes (MAST)\footnote{https://archive.stsci.edu/} %for 
to carry out photometric analysis of selected ApBp stars that are expected to possess a significant magnetic field. %In the process of first selection, the program G011060, written by Ernst Paunzen, was applied for preliminary selection ApBp candidates. 
%Since we are interested only in the ApBp stars that show rotational modulation in their light curves we have selected a sample of eight relatively bright objects ($V <8.0$~mag) with rotation periods longer than 2~d (see Table~\ref{tab1}).
We have selected a sample of eight relatively bright ApBp stars ($V <8.0$~mag) which show rotational modulation in their light curves with rotation periods longer than 2~d (see Table~\ref{tab1}).
%, the fraction of 10\% was distinguish from the sample. 

The selected ApBp stars possess relatively relatively low projected equatorial velocities ( $v\sin{i} < 65$ km~s$^{-1}$; see Table~\ref{tab1}) that may indicate the presence of a hydrodynamically stable atmosphere. Fast rotation leads to mixing in the stellar atmosphere that affects or even destroys abundance stratification in the absence of a magnetic field \citep{Michaud2015}. Therefore, we selected for this study %8
stars with rotation periods more than 2~d to ensure the hydrodynamic stability of their stellar atmosphere that is preferred for future analysis of abundance stratification. 
%Our reduction procedure does not affect frequency estimate of high-overtone pulsation in roAp star HD~24712.
Knowledge of precise values of rotation periods will help carry out abundance analysis of these stars using Magnetic Doppler Imaging.
%Since we are interested only in the stars with a long period of rotation, the list includes objects with the period $P> 2$ days, as well as bright objects $V <8.0$.

During the selection procedure we have used a database compiled by the TESS-AP procedure \citep{Khalack2019, Khalack2020} for stars observed by \textit{TESS} with 2~min cadence. This procedure automatically performs a Fourier analysis of the light curves by employing the code {\sc period04} \citep{Lenz2005} and collects stellar global parameters from the known astronomical databases (mainly from the \textit{TESS} Input Catalogue (TIC; \citealt{Stassun2018, Stassun2019}) and SIMBAD\footnote{http://simbad.u-strasbg.fr/simbad/}).
The sample is not complete and includes only 8 interesting targets (see Table~\ref{tab1}) selected to test our procedure for data reduction and analysis.
%Should emphasize that a significant part of the observed objects was 
Most of the selected targets have been observed in several sectors, which allowed us to effectively detrended their light curves and significantly reduce the instrumental distortions. %caused by \textit{TESS}'s CCD.
%to exclude the probability of distortion of data in the process of fitting them. 
%
%Please, describe here the reduction procedure!!!
%

%For the analysis of selected targets, we use the program TESS-AP \citep{Khalack2020, Khalack2019} for the Fourier transform of the light curve by employing the code {\sc period04} \citep{Lenz2005} and collecting stellar global parameters. Most of the global parameters were extracted from the MAST catalog with additional data provided by SIMBAD\footnote{http://simbad.u-strasbg.fr/simbad/} particularly for the values v$\sin{i}$ and v$_{\rm r}$. 

%Before starting the analysis, we focused on data preparation, to be specific the reduction of noise introduced by the instrument. This contribution is 

The instrumental distortions of a light curve are especially noticeable at the beginning and at the end of every 13.7~d segment of observations, corresponding to the \textit{TESS} orbital period (see, for example, top panel of Fig.~\ref{detrending}). %, ie the period of the orbit through the telescope. 

Since all the selected stars exhibit rotational modulation, the first step was to use the code {\sc period04} to detect and 
%extract the frequency of the rotational modulation using pre-whitening method,and its dominant harmonics (no more than three frequencies in total) from the light curve to obtain noisy data
extract the rotational frequency %of the rotational modulation 
and its dominant harmonics (no more than three frequencies in total) from the light curve using pre-whitening to obtain residuals with presumably only instrumental noise and distortions (see Fig.~\ref{detrending}, second panel from the top).
%The next step is to describe the noise with the help of a third-order polynomial. 
In general, the intervals for approximation were chosen as half the length of the observed sector or at the area between the gaps present in the light curve. 
It is worth emphasizing that at this stage one should choose
%it is worth choosing 
carefully the intervals used for the polynomial fits, because an insufficient approximation leaves a significant amount of distortion. Meanwhile, an over-correction using a high-order polynomial (more than 3) introduces additional distortion to the light curve, which in turn may lead to a different value of the period of stellar rotation. Then, we approximated the instrumental distortions by a third-order polynomial with the aim to remove them and obtain residuals with random noise only (see Fig.~\ref{detrending}, 3rd panel from the top).
As the final step, the same polynomials were subtracted from the initial light curve to remove the instrumental distortions and obtain a light curve only with the rotational modulation, the remaining noise (see bottom panel at the Fig.~\ref{detrending}), and also pulsation signals if present. 
A similar approach was used by \citet{Bowman2018} in the detrending of light curves of ApBp stars from the K2 mission.
%in reducing the contribution from the instrument is to extract this approximation from the initial data. 

%between the first interval of the light curve period. 
%We do not provide this image anymore!!!
%In the case when a sufficient number of $\langle B_{\rm z}\rangle$ measurements are available to determine the rotational period, the zero phase was set to the maximum of the mean longitudinal magnetic field which is closest to the first available measurement (see the case of HD~86592).  

%in the same manner on the initial interval period of polarimetric observations. 
We employed Monte Carlo simulations \citep{Bevington1969} with the code {\sc period04} to determine the precision of defined frequencies which results in %high precision
error bars of rotational periods.
Error bars for the derived phases of the magnetic field measurements include the error of the period estimation. % as well as the error of 1~min (5~min) of each measurement performed by \textit{TESS} with the 2~min (10~min) cadence which corresponds to 0.00069~d (0.00347~d). 
%In the phase error, we included the period error and the phase error implicated by the instrument, which is set to 0.00069 days. 
%For some stars, as HD24712 or HD74521, a significant contribution to the phase error-bars for magnetic observations was found due to a sharp difference in the time range of observations. 
%At the same time, the phase error bars are almost imperceptible considering a sufficiently high accuracy of estimation of the rotational period and recent spectropolarimetric observations that we used to measure the the mean longitudinal magnetic field.
%derived of the magnetic field. 
These error bars on the phase are small, given the tight constraints on the rotational periods and, in most cases, the relatively short intervals of time between the magnetic field measurements shown in the phase diagram plots and the time origin of the derived ephemerides.% (see Figs.~\ref{HD10840rot}, \ref{HD22920rot}, \ref{HD24712phase}, \ref{fig38170}, \ref{figHD63401}, \ref{figHD74521}, \ref{figHD77314} and \ref{figHD86592}).

\subsection{Spectroscopy}
\label{stokesIV}

For two stars (HD~22920 and HD~77314) we used unpolarized spectra obtained with the spectrograph HERMES (High-Efficiency and high-Resolution Mercator Echelle Spectrograph) at the Mercator telescope on La Palma.

The high-resolution fiber-fed prism-cross-dispersed echelle spectrograph HERMES covers the spectral domain from 3770 \AA\, to 9000 \AA\, in a single exposure with a resolution \emph{R} = 85000 \citep{Raskin2011}, which is suited for abundance analysis. It is bench-mounted and kept in an environment with the pressure and temperature control to assure the stability of its work. With the HERMES spectrograph, we obtained %Stokes I 
spectra of HD~22920 and HD~77314 with a relatively high signal-to-noise ratio (SNR $\sim$ 300). The spectra were reduced using the dedicated data reduction pipeline HermesDRS (version 6.0) \citep{Raskin2011}\footnote{For more details about this spectrograph and the reduction procedure, see {\rm  http://www.mercator.iac.es/instruments/hermes/}}, and used to derive effective temperature $T_{\rm eff}$, surface gravity $\log{g}$, radial velocity v$_{\rm r}$ and v$\sin{i}$ from analysis of Balmer line profiles (see Section~\ref{Balmer} and Table~\ref{tab1}).

\subsection{Spectropolarimetry}
\label{spectropolarimetry}

The spectra have been obtained with the spectropolarimeter ESPaDOnS (Echelle SpectroPolarimetric Device for Observations of Stars) at the Canada-France-Hawaii Telescope (CFHT) and used to derive global stellar parameters (see Section~\ref{Balmer}) and to measure the mean longitudinal magnetic field $\langle B_z\rangle$ (see Section~\ref{Bz_measurements}).

%, and with the spectrograph HERMES (High-Efficiency and high-Resolution Mercator Echelle Spectrograph) at the Mercator telescope on La Palma.
%The first and the second column in the Table~\ref{tab1} present respectively the spectrum identification of and the Barycentric Julian Date of observations in Barycentric Dynamical Time. The exposure time and the signal-to-noise ratio of the obtained spectra are shown is the third and forth columns, while the names of instruments used for the observations are specified in the fifth column. For the HERMES observations we provide the signal-to-noise ratio only for the Stokes I spectra.
%Employing the deep-depletion e2v device Olapa, 

%The spectropolarimeter 
ESPaDOnS\footnote{For more details about this instrument, see {\rm http://www.cfht.hawaii.edu/Instruments/Spectroscopy/Espadons/}} is capable of acquiring high resolution (R=65000) Stokes I and V spectra in the spectral domain from 3700 \AA, to 10000 \AA, with high SNR (in our case, spectra of targeted stars have SNR $ > 300$). The optical characteristics of 
%these twin
this spectrograph, as well as its performances, were described by \citet{Donati2006} and \citet{Wade2016}. The dedicated software package Libre-ESpRIT \citep{Donati1997} was employed to reduce the obtained Stokes I spectra and the Stokes V circular polarisation spectra as well.

Two selected targets, HD~77314 and HD~86592, were monitored with {\em dimaPol}, a  polarimeter module mounted at the entrance slit of the Cassegrain spectrograph installed on the 1.8-m Plaskett Telescope of the Dominion Astrophysical Observatory (DAO, \citealt{Monin2015}).  It is used to carry out circular spectropolarimetry of magnetic stars with resolution R$\sim$10000 in a 280\AA\, wide spectral region centered on the $H_{\beta}$ Balmer line.  By using Stokes V observations of the hydrogen $H_{\beta}$ line to measure the hemispherically averaged longitudinal magnetic field, {\em dimaPol} field measurements are %less effected by rotational broadening and is able to measure $\langle B_z\rangle$ in stars with relatively large values of v$\sin{i}$.
less affected by rotational broadening than measurements based on metal lines, so that they are better suited to measure $\langle B_z\rangle$ in stars with relatively large values of v$\sin{i}$. For a description of the instrument and details of the typical data acquisition and reduction procedures see \citet{Monin2015}.

\subsection{Analysis of Balmer line profiles}
\label{Balmer}

We used the available high-resolution spectra of the selected targets to estimate their effective temperature and surface gravity, and to measure their radial velocities and v$\sin{i}$ values (see for detail \citealt{Khalack+LeBlanc2015}). The Stokes I Balmer line profiles ($H_{\beta}$, $H_{\gamma}$, $H_{\delta}$, $H_{\epsilon}$, $H8$, $H9$, $H10$, $H11$, and sometimes $H12$) and the lines of metals located in the wings of Balmer lines were fitted by the theoretical profiles with the help of {\sc fitsb2} code \citep{Napiwotzki2004} employing the grids of stellar atmosphere models \citep{Husser2013} calculated using the {\sc phoenix-16} code \citep{Hauschildt+Baron1999}. The derived values of the global stellar parameters mostly fall into the range of estimation error bars compared to the previously published data (see Table~\ref{tab1}) and to the results obtained in this study from the Geneva and Str{\"o}mgren-Crawford photometry (see Table~\ref{Geneva}).

\subsection{Measurements of the %hemispherically averaged 
mean longitudinal magnetic field.}
\label{Bz_measurements}

For each star, the measurements of the hemispherically averaged longitudinal magnetic field $\langle B_{\rm z}\rangle$ were extracted from the literature or derived from the analysis of available spectropolarimetric observations, and are presented in Table~\ref{appendix} and discussed in Section~\ref{results}. 
The $\langle B_{\rm z}\rangle$ values  have been derived from analysis of spectra obtained with different instruments \citep{Donati2006, Bagnulo2015, Monin2015} and using different methods \citep{Donati1997_2, Bagnulo2002, Kochukhov2010, Mathys2017}, and may be inconsistent with each other. This information should be considered during analysis of $\langle B_{\rm z}\rangle$ variability with rotational phase.

We determined the mean longitudinal magnetic field $\langle B_z\rangle$ of HD~24712, HD~63401 and HD~74521 at the various epochs of observation by application of the moment technique to the Stokes I and V spectra of these stars that were observed with ESPaDOnS. The analysis procedure is as described by \citet{Mathys2017}. In short, in each spectrum, we measured the first order moments about the central wavelengths $\lambda_0$ of the Stokes V profiles of a set of selected diagnostic lines and performed a least-squares fit to derive the value of $\langle B_z\rangle$ from the slope of the linear dependence of these moments on the product of the effective Land\'e factors of the corresponding transitions by $\lambda_0^2$. We used the standard error of the longitudinal field that is derived from this least-squares analysis as an estimate of the uncertainty affecting the obtained value of $\langle B_z\rangle$. For HD~24712, we used Fe~{\sc i} lines as diagnostic lines, and for HD~74521, we employed Fe~{\sc ii} lines. Lines of these two ions were systematically used by \citet{Mathys2017} for magnetic field determinations, on account of the fact that inhomogeneities in the distribution of iron over the surfaces of Ap stars tend to be moderate, so that  measurements based on the analysis of Fe lines are mostly representative of the intensity and structure of the magnetic field itself. However, for HD~63401, whose $v\,\sin i$ is higher than for HD~24712 or HD~74521, the Fe lines proved ill-suited for magnetic field determination, so we used Si~{\sc ii} lines instead as diagnostic lines. Due to the highly non-uniform distribution of Si on the surface of HD~63401, these lines show considerable distortions, variable with rotation phase,  which complicate their measurements. One should keep in mind that the derived $\langle B_z\rangle$ values are representative of the %common
convolved contribution of the geometrical structure of the magnetic field and of the inhomogeneous distribution of Si.

%For HD~77314 and HD~86592, the reported $\langle B_z\rangle$ measurements (see Table~\ref{appendix}) have been obtained with the help of dimaPol spectropolarimeter \citep{Monin2015}. This is a  Cassegrain spectrograph installed on 1.8-m Plaskett telescope of the Dominion Astrophysical Observatory (DAO). It is used to carry out circular spectropolarimetry of magnetic stars with resolution R$\sim$10000 in the 280\AA\, wide spectral region centered on the $H_{\beta}$ Balmer line. The spectropolarimeter dimaPol employs hydrogen $H_{\beta}$ and metallic lines around it to measure the hemispherically averaged longitudinal magnetic field, and therefore is less effected by rotational broadening and is capable to measure $\langle B_z\rangle$ in stars with relatively large values of v$\sin{i}$ (see for more details \citealt{Monin2015}).

\begin{table*}
\begin{center}
\caption{Stellar parameters derived from Str{\"o}mgren-Crawford photometry ($T_{\rm eff}$ in column~2), from BC (log($L_{\star}$ in column~4), and from isochrone fitting for the studied stars (columns~3, 6, 8, 10). Data for stellar luminosity, radius and mass inferred from TIC~8  \citep{Stassun2019} are shown for comparison in column~5, 7, 9 respectively. Data for the derived equatorial velocity are presented in the last two columns.} %The first column for each star presents values derived by the method that is described in the Discussion section, the second column contains values extracted from TESS catalog TIC~8}
\label{Geneva}
\def\arraystretch{0.98}
\setlength{\tabcolsep}{3pt}
\begin{tabular}{lccccccccccc}
\hline
Name & $T_{\rm eff}$ (K) & $\log{g}$ & \multicolumn{2}{c}{log($L_{\star}$/ $L_{\sun}$)}& \multicolumn{2}{c}{$R$ ($R_{\sun}$)}& \multicolumn{2}{c}{$M_{\star}$ ($M_{\sun}$)} &  logAge & $v_{\rm eq}$  (km~s$^{-1}$) & $v_{\rm eq}/v_{\rm crit}$ ($\%$)\\

HD & This study & This study & This study & TIC & This study & TIC & This study & TIC & This study & This study & This study\\
& Photometry & Isochrone & BC cor. & & Isochrone & & Isochrone & &Isochrone & & \\ 
\hline
10840 & 11489 $\pm$ 316 & 4.13(2)& 1.97(3) & 2.035(35)& 2.452(3) & 2.550(80) & 2.923(1) & 3.09(39)& 8.224(3) &  59.12(7) & 12.40(2) \\
22920 & 13714 $\pm$ 386 & 3.91(2) & 2.65(3)& 2.424(320)& 3.776(4) & - & 4.152(1) & - & 8.0734(4) &  48.39(5) & 10.57(2) \\
24712 & 7201 $\pm$ 290 & 4.16(6) & 0.85(1) & 0.863(15) & 1.724(2) & 1.715(70) & 1.545(1) &  1.63(27)& 9.03(6) &  7.00(1) & 1.69(1) \\
38170 & 9493 $\pm$ 293 & 3.79(2) &1.94(3) & 2.006(34) & 3.463(3) & 3.354(142) & 2.6762(7) & 2.56(35) & 8.6324(3) &   63.32(5) & 16.50(2) \\
63401 & 13201 $\pm$ 380 & 4.15(2) & 2.27(3) & - & 2.617(3) & - & 3.526(1) & - & 7.946(2) &  54.82(2) & 10.82(2)\\
74521 & 10474 $\pm$ 285 & 3.80(2) & 2.16(3) & - & 3.657(3) & - & 3.034(1) & - & 8.4811(3) &  26.24(2) & 6.60(1)\\
86592 & 7716 $\pm$ 336 & 4.08(6) & 1.10(1) & 1.150(23) & 2.003(3) & 1.895(60) & 1.7482(5) & 1.981(300) & 8.986(2)&  35.10(5) & 8.61(2) \\
\hline

\end{tabular}
\end{center}
\end{table*}

For HD~77314 and HD~86592, the reported $\langle B_z\rangle$ measurements (see Table~\ref{appendix}) have been obtained with {\em dimaPol} \citep{Monin2015}, which are less affected by considerable rotational broadening of this targets due to the designed qualities of this unique instrument (see Section~\ref{spectropolarimetry}).%, a  polarimeter module mounted at the entrance slit of the Cassegrain spectrograph installed on the 1.8-m Plaskett Telescope of the Dominion Astrophysical Observatory (DAO, \citealt{Monin2015}).
%It is used to carry out circular spectropolarimetry of magnetic stars with resolution R$\sim$10000 in a 280\AA\, wide spectral region centered on the $H_{\beta}$ Balmer line.  By using Stokes V observations of the hydrogen $H_{\beta}$ line to measure the hemispherically averaged longitudinal magnetic field, {\em dimaPol} field measurements are %less effected by rotational broadening and is able to measure $\langle B_z\rangle$ in stars with relatively large values of v$\sin{i}$.
%less affected by rotational broadening than measurements based on metal lines, so that they are better suited to measure $\langle B_z\rangle$ in stars with relatively large values of v$\sin{i}$. For a description of the instrument and details of the typical data acquisition and reduction procedures see \citet{Monin2015}.}

The timestamps of flux and $\langle B_{\rm z}\rangle$ measurements were converted to the Barycentric Julian Date to use the same time units to plot the light curve and magnetic field phase diagrams. % with an idea to compare for each object the phase diagrams that correspond to stellar rotation and are built using the light curve and the magnetic field measurements. 
Since the rotation periods obtained from photometric observations were used to build a phase diagram for most stars, the zero phase for the magnetic and photometric data was set at the minimum closest to the beginning of the analysed light curve.
%of the light curve at the first interval of the observed period.

\subsection{Geneva and Str{\"o}mgren-Crawford photometry}

%In the light of the fact that global stellar parameters determined with different methods should have the same values, we used an alternative approach to assess them. 
Geneva and Str{\"o}mgren-Crawford photometric indices were used to determine the effective temperature as the first step in evaluation of the luminosity by using the method of bolometric correction (BC) \citep{Flower1996}, and radius through stellar isochrone fitting \citep{Sichevskij2017}. The derived stellar parameters (see Table~\ref{Geneva}) were employed to determine the rotational rate $v_{\rm eq}/v_{\rm crit}$ for the stars described in this study. 

For the Geneva photometry, the General Catalogue of Photometric  Data\footnote{https://gcpd.physics.muni.cz/} by \citet{Paunzen2015} was used. The values of $c1$, $m1$, and $(b-y)$ for the uvby$\beta$ photometry were taken from the \citet{Hauck1998} catalogue. Unfortunately, one star (HD77314) was left out of consideration because there are no data available for this object. To determine the value of $[u-b]$, we used the original formulae described by \citet{Stromgren1966A} as $[u-b] = [c1] + 2[m1]$, where reddening free indices are calculated from $[m1] = m1 + 0.18(b-y)$ and $[c1] = c1 - 0.20(b-y)$. Then, we have used the formulae for variable $\theta_{\rm eff}$ specified for CP2 stars %\citep{Preston1974}
to calibrate the effective temperature
\begin{equation}
    \theta_{\rm eff} = 0.234(0.009) + 0.213(0.008)[u-b]~,
\end{equation}
 \citet{Netopil2008} (for the uvby$\beta$ photometry) and 
 \begin{equation}
      \theta_{\rm eff} = 0.835(0.028) + 0.458(0.013)(B2-G)^{0}~,
 \end{equation}
 \citep{Hauck1982} (for the Geneva photometry). \citet{Rufener1988} reported that the standard deviation of colors in the Geneva photometry is around 0.001 mag. The final values of effective temperature are obtained from $\theta_{\rm eff} = 5040/T_{\rm eff}$ \citep{Napiwotzki1993}.

%A convenient way to estimate stellar luminosity is the bolometric correction (BC). 
The value of BC is calculated taking into account the coefficients from \citet{Flower1996} and the effective temperature obtained from photometry. %, the value of BC was calculated. 
To derive stellar luminosity $\log(L/L_{\sun})$:
\begin{equation}
    \log(L/L_{\sun}) = -0.4[M_{V} - V_{\sun} - 31.572 + (BC_{V} - BC_{V, \sun})]
\end{equation}
 we have employed the $BC_{V, \sun}$ = $-0.08$ and the apparent magnitude of the Sun $V_{\sun}$ = $-26.75\pm0.03$ which
 %are widely used by scientific community and
 are introduced by \citet{Cox2000} and later adjusted by \citet{Torres2010}. To determine the absolute magnitude $M_{V}$, we have used values of the derived parallax from Gaia EDR3\footnote{https://gea.esac.esa.int/archive/} and apparent magnitude in V band from the Hipparcos catalogue\footnote{https://hipparcos-tools.cosmos.esa.int/HIPcatalogueSearch.html}. 

The Stellar Isochrone Fitting Tool\footnote{https://github.com/Johaney-s/StIFT}
%developed by Johana Sup{\'i}kov{\'a}
was used to estimate stellar age, radius and mass with the methods described by \citet{Sichevskij2017}. Parameters were determined from evolutionary track models of solar metallicity $Z = 0.014$ \citep{Bressan2012}, according to the given temperature and luminosity. Using the following relation for the equatorial velocity \citep{Netopil2017}, where the stellar radius is expressed in the units of solar radius $R_{\sun}$,%, based on well known value of solar radius $R_{\sun}$
\begin{equation}
    v_{\rm eq} ({\rm km s^{-1}}) = 50.579R(R_{\sun})/P_{\rm rot}({\rm d})
\end{equation}
 we are able to derive the critical rotational fraction $v_{\rm eq}/v_{\rm crit}$. Considering the formula provided for critical velocity $v_{\rm crit}$ by \citet{Georgy2013}, we have derived the following expression for the rotation rate: 
\begin{equation} \label{veqvcrit}
v_{\rm eq}/v_{\rm crit} = \frac{50.579 R_{\rm eq}(R_{\sun})}{P_{\rm rot}}\sqrt{\frac{R_{\rm eq}(R_{\sun})}{G M_{\star}(M_{\sun})}}
\end{equation}
%is valid for objects having smaller masses than 5$M_{\sun}$, which fully satisfies our sample of the selected targets. For the magnetic CP stars, their rotational rate is usually below 0.5 (ref??)
%which is in a good agreement with our data obtained for the studied stars considering their rotational periods derived in this study (see Table~\ref{Discussion}).  %confirmed recently study for HgMn stars \citep{Kochukhov2021} and completely goes along with top 17$\%$ for selected stars. 
%Similar results for the rotational rates of HgMn stars have been obtained recently by \citet{Kochukhov2021}. 
where equatorial radius $R_{\rm eq}(R_{\sun})$, stellar mass $M_{\star}(M_{\sun})$ are expressed in respective solar units, and $G$ stands for the universal gravitational constant.

Summarising the procedure described above we have collected the data in Table~\ref{Geneva} and compared them to the values extracted from the TIC~8 catalogue \citep{Stassun2018, Stassun2019}. The age derived for each target is measured in years (see column 10 of Table~\ref{Geneva}).
%Change the next phrase and and explain ir please
%Even though the data slightly diverge between themself, c
The combination of the stellar parameters derived from the analysis of Balmer line profiles, from Geneva and Str{\"o}mgren-Crawford photometry (see Table~\ref{Geneva}), and from the literature (see Table~\ref{tab1}) in most cases points to the same % on average
values considering estimation error bars. Three stars, HD~74521, HD~86592, and HD~10840 are exceptions. HD~74521 shows a large discrepancy between $\log{g}$ determined from isochrone fitting and the TIC, and for $T_{\rm eff}$ in values extracted from the TIC while values derived from Balmer lines and Str{\"o}mgren-Crawford photometry matched. %Similar 
Discrepancy occurred also to the values $R$ and log($L_{\star}$/ $L_{\sun}$) considering their error-bars obtained for stars HD~10840 and HD~86592 from the TIC and from the isochrone fitting. %exceeded the uncertainties determined for $R$ and log($L_{\star}$/ $L_{\sun}$) respectively taking into account the values derived from isochrone fitting and TIC data.

\section{Results of analysis of individual stars}
\label{results}

\begin{figure}
\centering
\includegraphics[width=\linewidth]{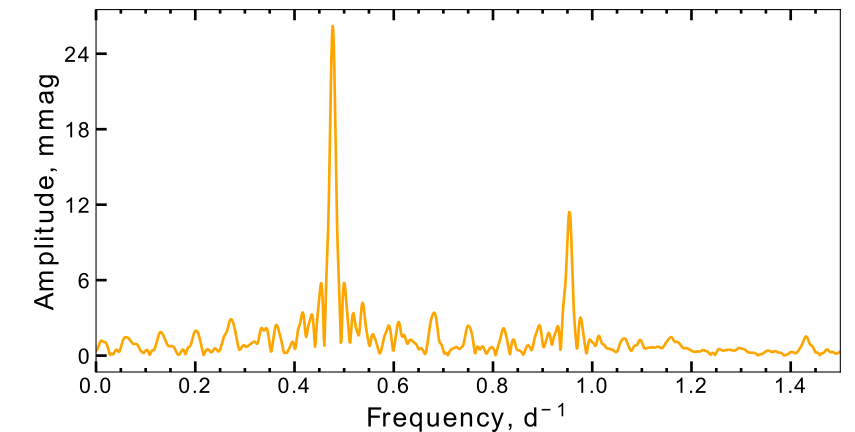}
\includegraphics[width=\linewidth]{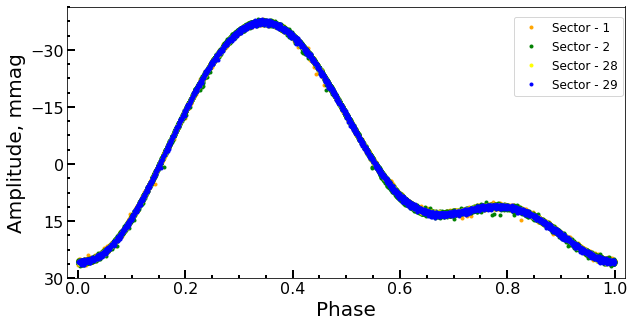}
\caption{Periodogram (top panel) and phase diagram (bottom panel) built for the rotational period $\emph{P} = 2.0976858$~d in HD~10840 using the reduced 2~min cadence light curve from the sectors~1 - 2 and 10~min cadence light curve from the sectors~28-29.}
%28-29 with 10~min cadence.
%and 10-minutes cadence observations in sectors 28-29. 
 %The $\langle B_{\rm z}\rangle$ measurement is taken from \citet{Bagnulo2015} (grey triangle). In Figure 2 (and all similar figures) there are references to the top and bottom panels, but no reference or description of the middle panel.
\label{HD10840rot}
\end{figure}

\begin{figure}
\centering
\includegraphics[width=\linewidth]{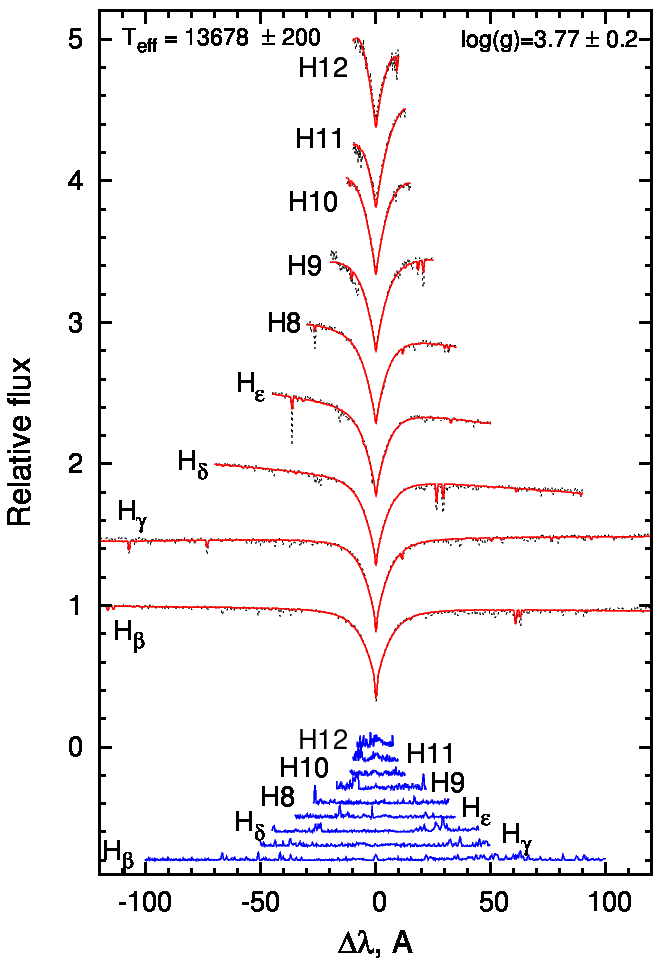}
\caption{
%Change it. It should be another spectrum shown here, with another results. 
The synthetic spectra (thin dotted lines) are fitting well the Balmer line profiles (thick lines) observed  by NARVAL for HD~22920 assuming %according to the values of 
$T_{\rm eff} = 13700 \pm 200$~K and $\log{g} = 3.8 \pm 0.2$ and metallicity $M = -0.5$. For the purpose of visual convenience, Balmer lines are shifted by 0.5. Difference between the synthetic and observed spectra is shown by blue lines below. }
\label{figHD22920Bal}
\end{figure}

\subsection{HD~10840 (TIC~231844926 = BM~Hyi)}\label{HD10840}

According to the catalog of \citet{Houk1975} HD~10840 is classified as a CP star of ApSi type and is on the boundary between the spectral types A and B
taking into account its effective temperature $T_{\rm eff} = 11600$~K and surface gravity $\log{g} = 3.60$ \citep{Bailey2013}. %Therefore, 
\citet{Renson1992} provided for this star a spectral type of B9.
\citet{Levato1996} derived radial velocity v$_{\rm r} = 19.4\pm 2.1$~km~s$^{-1}$ and v$\sin{i} < 30$~km~s$^{-1}$.
%basing on the Hipparcos satellite observations. 
\citet{Bailey2013} %used the code {\sc zeeman} to carry out an abundance analysis of HD~10840's spectra, 
determined its v$\sin{i}$ = 35$\pm$5~km~s$^{-1}$.
%, and found significantly overabundant Ti, Cr, Fe, and Si in its atmosphere. 
%As well as have used code {\sc zeeman} to carry out an abundance analysis of Ti, Cr, Fe, and Si, and found these elements significantly overabundant in the stellar atmosphere of HD~10840.

Measurement of the mean longitudinal magnetic field of HD~10840 were obtained by \citet{Kochukhov2006} and \citet{Bagnulo2015}. % (see Table~\ref{HD10840_Bz}).
\citet{Kochukhov2006} used the method developed by \citet{Bagnulo2002} to derive the mean longitudinal magnetic field from analysis of circular polarisation detected in FORS1 spectra in the area of the Balmer line profiles.
\citet{Bagnulo2015} employed the same method to analyze the same FORS1 spectrum of HD~10840, but taking into account the polarisation detected in the Balmer and metal line profiles (see Table~\ref{appendix}).
%line profiles and in the line profiles of helium and 
%other metals (see Table~\ref{appendix}).
\citet{Bailey2013} used a simple magnetic dipole model with the adopted field strength at the magnetic pole, $B_{\rm d}$=500~G, to fit the magnetically sensitive line profiles of Ti, Cr, Fe and Si found in the available spectra of HD~10840. 
%measurements of magnetic stars to describe the magnetic field structure of HD~10840 in the frame of a simple magnetic dipole model and estimated the field strength at the magnetic pole, $B_{\rm d}$=500~G.

\begin{figure}
\centering
\includegraphics[width=\linewidth]{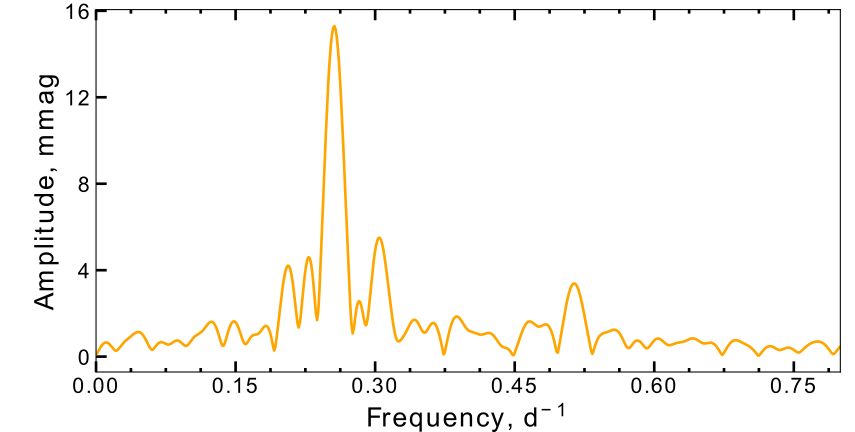}
\includegraphics[width=\linewidth]{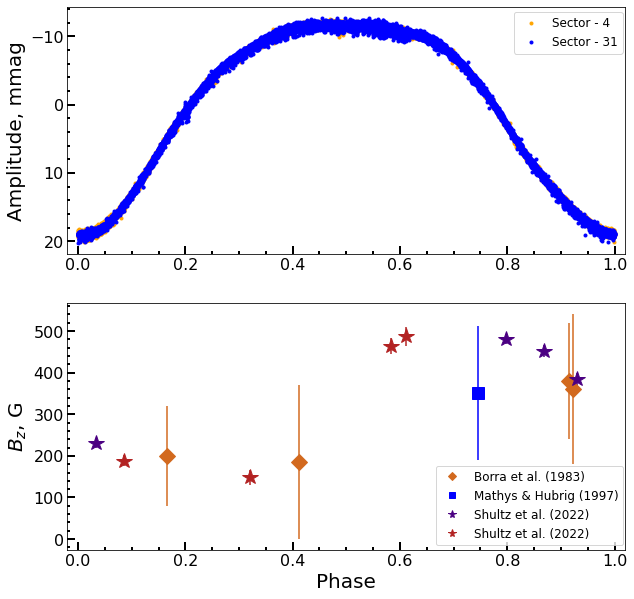}
\caption{%The same as at the Fig~\ref{HD10840rot} but for HD~22920. 
Periodogram (top panel) and phase diagram (middle panel) built for rotational period 3.947225~d of HD~22920 using the reduced light curve obtained from sectors~4 and 31. The bottom panel shows phase diagram built on the measurements of the mean longitudinal magnetic field taken from \citet{Borra1983} (brown diamonds), \citet{Mathys1997} (blue squares). We show also  the LSD measurements of the mean longitudinal magnetic field obtained by \citet{Shultz2022} from NARVAL (red stars) and ESPaDOnS (indigo stars) spectra.}
\label{HD22920rot}
\end{figure}

\begin{figure}
\centering
\includegraphics[width=\linewidth]{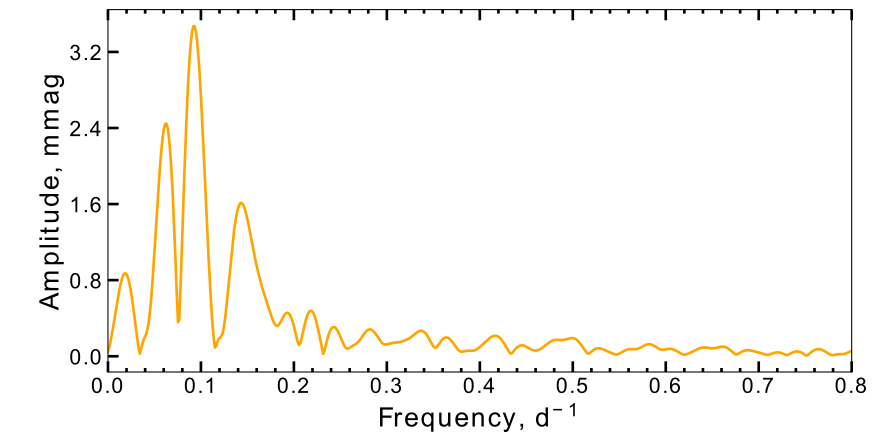}
\includegraphics[width=\linewidth]{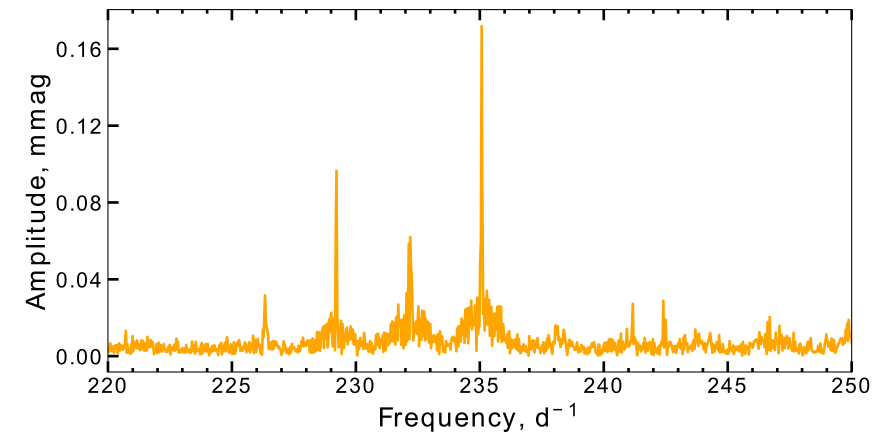}
\caption{Discrete Fourier transform of the reduced light curve obtained by TESS for HD~24712 in the sector 5 and 31.  Significant signals are detected at low frequencies (top panel) that correspond to stellar rotation, and at the high frequencies (bottom panel) that correspond to roAp pulsations. }
\label{HD24712}
\end{figure}

The period $\emph{P} = 2.097679(7)$~d of stellar rotation of HD~10840 was determined by \citet{Sikora2019} from the analysis of
photometric data obtained with the space telescope \textit{TESS} in sectors~1-2. In our study, we measure a more precise value of the rotation period as $\emph{P} = 2.0976858(2)$~d which is compatible with that from \citet{Sikora2019}. It was derived from the analysis of detrended light curves obtained for this star in sectors 1 and 2 with 2~min cadence and in sectors 28 and 29 with 10~min cadence.

%and provide the respective phase diagram for the variability of light curve with stellar rotation (see the lower panel in Fig.~\ref{HD10840rot}).
The upper panel of Fig.~\ref{HD10840rot} presents the low-frequency region of the discrete Fourier transform of the reduced light curve of HD~10840. 
From the Fourier analysis we identified a signal at $\nu = 0.47671583(5)$~d$^{-1}$ and its first harmonic.
%We assume that the detected signal is generated by the spotted abundance distribution of chemical elements in the stellar atmosphere of HD~10840 that causes a flux modulation with the stellar rotation. 
In this case, the derived lower frequency, which appears to be the tallest peak, is identified as the frequency of stellar rotation.% and on contrary to the case of the star HD~38170 (see Section~\ref{HD38170text}).
The lower panel of Fig.~\ref{HD10840rot} shows the phase diagram built using the derived rotational period.% and the rotational phase of the available $\langle B_{\rm z}\rangle$ measurement.
%variability in the mean longitudinal magnetic field ${B}_{\rm z}$ with the phase of stellar rotation.
%The available measurements of ${B}_{\rm z}$  (see Tab.~\ref{HD10840_Bz}) is located close to the phases of the best visibility of magnetic poles if one assumes the OMR model.

\subsection{HD~22920 (TIC~301621458 = FY~Eri)}

%Initially, HD~22920 was reported by \citet{Cowley1972} as a variable CP star of spectral type B9~IIIp~Si. Later it was reclassified  %although in the latest studies spectral type was established 
%as B8 \citep{Houk1999} that corresponds to the $T_{\rm eff}$ = 13700~K and $\log{g}$ = 3.72 found by \citet{Catanzaro+1999}.

\begin{figure}
\centering
\includegraphics[width=\linewidth]{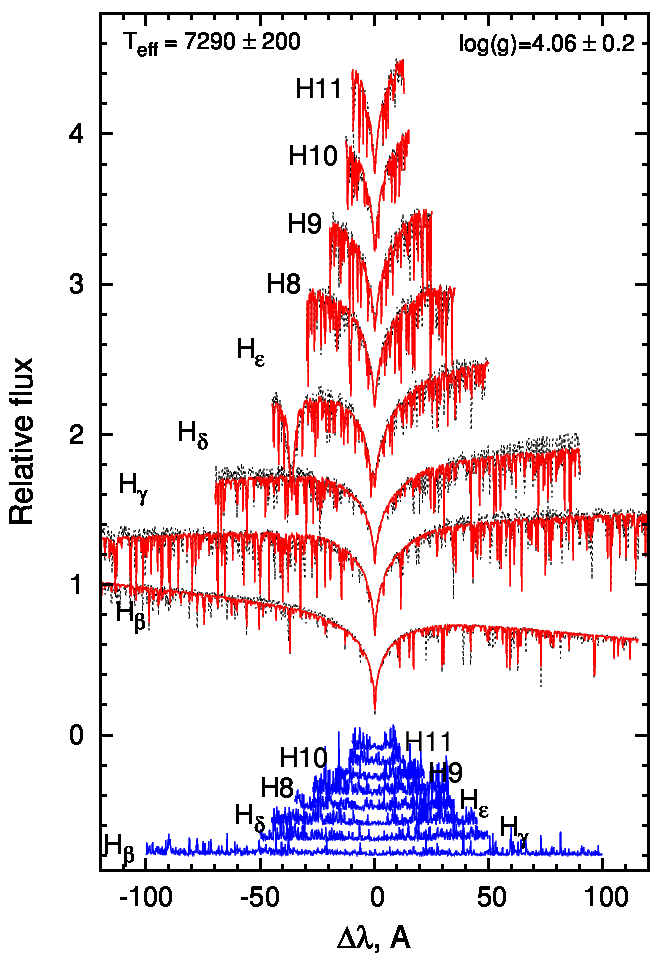}
\caption{The same as in Fig.~\ref{figHD22920Bal} but the best fit of Balmer line profiles in HD~24712 is obtained for %except for the synthetic lines of the star 
 $T_{\rm eff} = 7300 \pm 200$~K and $\log{g} = 4.1 \pm 0.2$. }
\label{HD24712Balmer}
\end{figure}

\begin{figure}
\centering
\includegraphics[width=\linewidth]{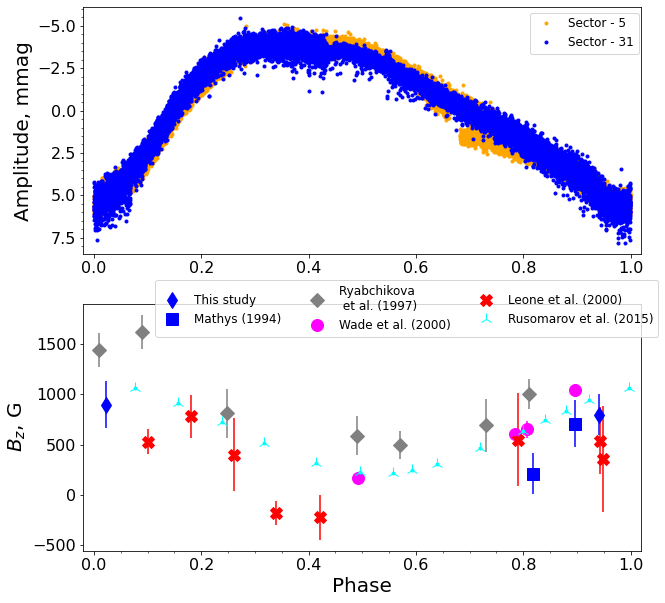}
\caption{
Phase diagram of HD~24712 is built using the derived rotational period $\emph{P} = 12.24267$~d for the reduced light curve from the sector 5 and 31, and for the $\langle B_{\rm z}\rangle$ measurement from  \citet{Wade2000} (pink circles), \citet{Mathys1994} (blue squares), and \citet{Leone2000} (red cross).% obtained for HD~24712.
}
\label{HD24712phase}
\end{figure}

HD~22920 was reported by \citet{Cowley1968} as a variable CP star of spectral type B8pSi, and the most recently determined spectral type is B8II \citep{Houk1999}. \citet{Leone+Manfre1996} found for this star $T_{\rm eff} = 13800$~K and $\log{g} = 3.65$, which are close to the estimates of $T_{\rm eff} = 13700$~K and $\log{g} = 3.72$ obtained by \citet{Catanzaro+1999} and to the results $T_{\rm eff} = 13640$~K and $\log{g} = 3.74$ derived by \citet{Khalack+LeBlanc2015}. From the recently obtained ESPaDOnS spectropolarimetric data, we measured the values of $T_{\rm eff}= 13700 \pm 200$~K and $\log{g} = 3.8 \pm 0.2$ from fitting Balmer line profiles with the help of the {\sc fitsb2} code \citep{Napiwotzki2004} %and using the the grid of fluxes calculated with {\sc phoenix-16} code \citep{Husser2013} 
(see Fig.~\ref{figHD22920Bal}).

%\citet{Khalack2015} analyzed the spectra acquired with ESPaDOnS to study variability of element's abundance with the rotational phase of HD~22920. These authors reported that line profiles of oxygen, silicon, iron and chromium
%change with the rotational phase suggesting the idea that their abundance is horizontally stratified in the stellar atmosphere %of non-uniform horizontal distributions 
%of HD~22920. The abundance of silicon and iron also was found to be vertically stratified \citep{Khalack2015}.
From the analysis of nine Balmer line profiles, \citet{Khalack2015} estimated $T_{\rm eff}$= 13640~K and $\log{g}$ = 3.72. The LTE model of the stellar atmosphere has been calculated by \citet{Khalack2015} employing the code PHOENIX \citep{Hauschildt1997} for the derived effective temperature and $\log{g}$ to carry out the line profile simulations with the help of the {\sc zeeman} code \citep{Landstreet1982}. This analysis demonstrated that HD~22920 has inhomogeneous abundance distribution of chemical elements in its atmosphere considering that it shows a variability of line profiles for the studied ions.

%The values of v$_{\rm r}$=18.0~km~s$^{-1}$, v$\sin{i}$ = 37.0~km~s$^{-1}$ (presented in Table~\ref{tab1}) and abundance (oxygen, silicon, iron, chromium) have been specified using an automatic minimization routine that treats individually each profile line .

%The intensity $Z_0=-0.014$ of the  $\uplambda5200$\r{A} depression has been used to estimate the metalicity in this star. Basing on the large statistical probing of $Z_0$, a presence of the weak chemical anomalies and weak magnetic field was determined for HD~22920 \citep{Glagolevskij2007}.

\citet{Mathys1997} and \citet{Borra1983} analysed 
spectropolarimetric data obtained by using the Zeeman analyzer at Cassegrain Echelle Spectrograph (CASPEC) and photoelectric Pockels cell polarimeter respectively, %therefore, to 
and derived the mean longitudinal magnetic field in the different rotational phases of HD~22920 (see Table~\ref{appendix}). \citet{Glagolevskij2007} calculated the  root-mean-square magnetic field (line-of-sight component) $<B_{\rm e}> = 148 \pm 50$~G and average surface magnetic field $B_{\rm s}= 800$~G.
Applying the least-squares deconvolution (LSD) method \citep{Donati1997_2, Kochukhov2010} to the available ESPaDOnS and NARVAL spectra of HD~22920, \citet{Shultz2022} measured the mean longitudinal magnetic field (see Table~\ref{appendix}) for an additional eight rotational phases and provided a precise value for the dipolar magnetic field $B_{\rm d} = 1.6^{+1.1}_{-0.0}$~kG.

%\citet{Glagolevskij2007} studied HD~22920 to investigate the relationship between the derived abundance and the weak magnetic field.  The author calculated the  root-mean-square magnetic field (line-of-sight component) $<B_{\rm e}>$ =148$\pm$50~G and average surface magnetic field $B_{\rm s}$= 800~G.

%, phase diagram calculate of , and mean longitudinal magnetic field measured for HD~22920. 
%Considering that the detected signal on the discrete Fourier transform most probably was caused by the existence of spots of chemical anomalies in the stellar atmosphere, we assume that it corresponds to the period of stellar rotation.
\citet{Bartholdi1988} found that this star shows a periodic variability in Geneva 7-color photometry with $\emph{P} = 3.95$~d. %This period is slightly different from the recently published period
The value of this period was recently refined to $\emph{P} = 3.9489(3)$~d from the MASCARA photometric survey \citep{Bernhard2020}, and to the period $\emph{P} = 3.9472(1)$~d found by \citet{Shultz2022} via the fitting of $<B_{\rm z}>$ measurements.
Based on our detrended light curve from sectors 4 and 31 we have determined the value of the rotation period to be  $\emph{P} = 3.947225(2)$~d for HD~22920 which is in good accordance with the result published by \citet{Shultz2022}. The top panel of Fig.~\ref{HD22920rot} shows the discrete Fourier transform of the reduced light curve at low frequencies for HD~22920. 
We have used the derived period to build a phase diagram of the light curve (see middle panel of Fig.~\ref{HD22920rot}) and magnetic field measurements (see bottom panel of Fig.~\ref{HD22920rot}). %which gives best correlation between available data. %even though 
%Using the derived data we try to estimate the correlation of the magnetic field with the rotation of HD~22920. 
One can see from the bottom panel of Fig.~\ref{HD22920rot} that the mean longitudinal magnetic field reaches its minimum at the phase $\varphi \simeq$ 0.3 when the light curve is close to its maximum, and increases to maximum field at the phase  $\varphi \simeq$ 0.7 while the light curve is decreasing.  
It appears that the magnetic field influences the %horizontal stratification
departure from horizontal uniformity of the chemical composition in stellar atmosphere of HD~22920, which in turn causes the observed rotational modulation of the light curve assuming the oblique rotator model.

\subsection{HD~24712 (TIC~279485093 = DO~Eri)}

\begin{figure}
\centering
\includegraphics[width=\linewidth]{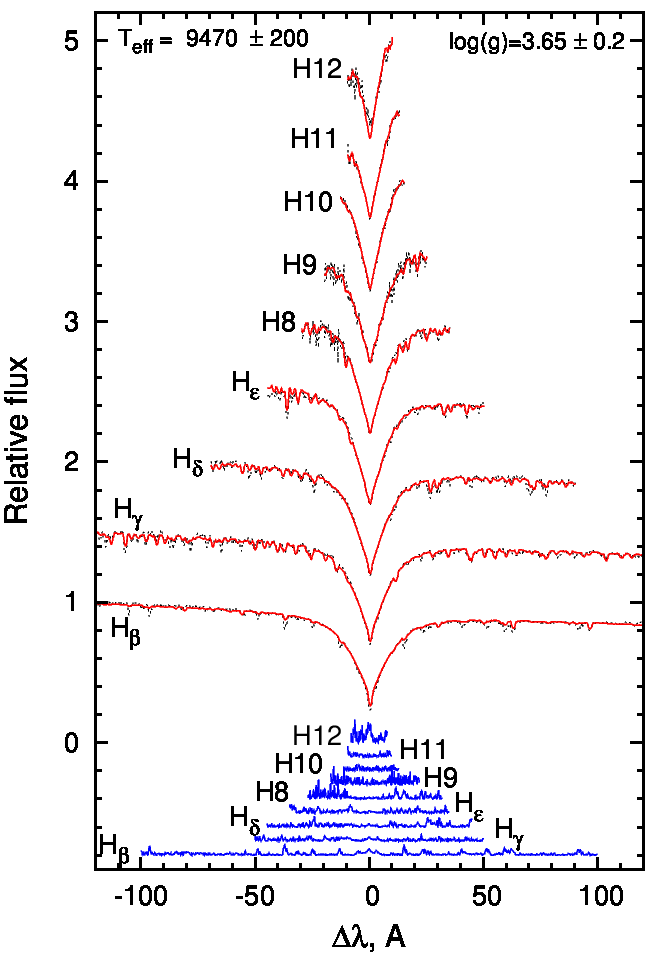}
\caption{The same type of analysis as in Fig.~\ref{figHD22920Bal} but the best fit of Balmer lines in HD~38170 is obtained  %except for the synthetic lines of the star HD~38170 were used the value of 
for $T_{\rm eff} = 9500 \pm 200$~K and $\log{g} = 3.65 \pm 0.2$. }
\label{HD38170Balmer}
\end{figure}

\begin{figure}
\centering
\includegraphics[width=\linewidth]{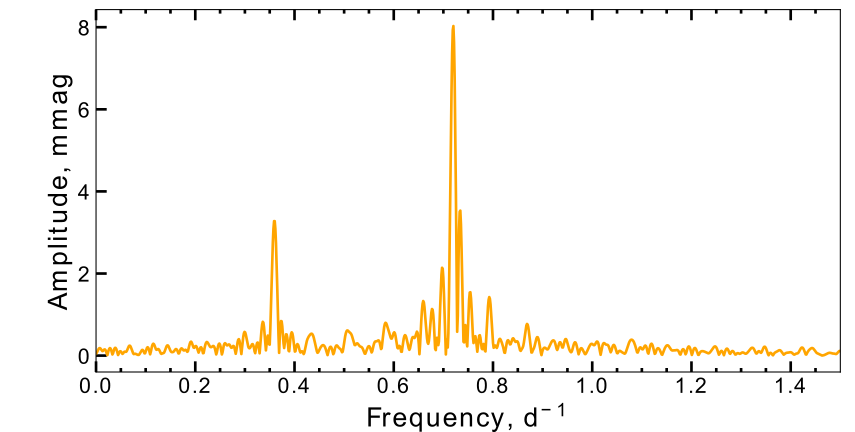}
\includegraphics[width=\linewidth]{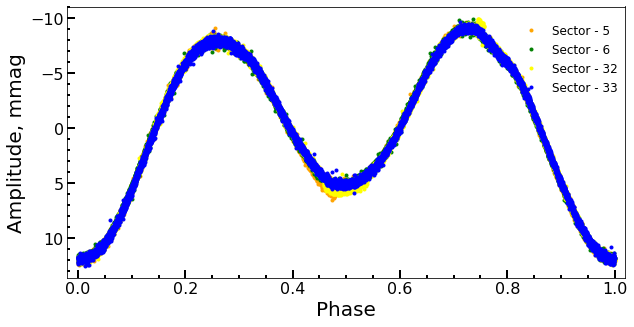}
\caption{The same as at the Fig~\ref{HD10840rot} but for HD~38170. The phase diagram is built using the rotational period $\emph{P} = 2.766116$~d derived from the reduced light curve observed in the sector~5, 6, 32 and 33.% for the $\langle B_{\rm z}\rangle$ measurement from \citet{David-Uraz2021} (pink triangle).
}
\label{fig38170}
\end{figure}

HD~24712 is one of the best-studied rapidly oscillating 
(roAp) chemically peculiar stars.
%From the study of the relation between radial velocity and spectral peculiarities 
\citet{Abt1995} determined that HD~24712 is a variable star of spectral type A9Vp SrEuCr. \citet{Preston1972} measured for this star the values of v$\sin{i} = 7.0$~km~s$^{-1}$ and $i = 40^\circ$. % After Fourier analysis of several Balmer line profiles, \citet{Royer2002} derived v$\sin{i} = 18.0$~km~s$^{-1}$ from the first harmonic.
In the study of mCP stars \citet{Sikora2019_2} derived measurements of v$\sin{i} = 6.6 \pm 0.6$~km~s$^{-1}$ and $\log{g} = 3.8 \pm 1.0$ from spectroscopy.
%In the profound investigation of this specific star, 
\citet{Ryabchikova1997} provide the values of $T_{\rm eff} = 7250$~K and $\log{g} = 4.2$ from the photometric and spectroscopic observations as well as magnetic measurements. Using available spectroscopic data we determine %the improved values 
 $T_{\rm eff} = 7300 \pm 200$~K, $\log{g} = 4.1 \pm 0.2$ %and v$\sin{i} = 22 \pm 2$~km~s$^{-1}$ 
 from the analysis of Balmer line profiles (see Fig.~\ref{HD24712Balmer}) which are in a good accordance with the previously published data considering the estimated error bars.

Assuming the oblique rotator model, \citet{Preston1972} derived a rotation period of $\emph{P} = 12.448$~d from the EWs of metal lines.
%and described in the term of OMR model. 
 Later, \citet{Kurtz1987} improved the estimation of the photometric rotational period to 12.4572(3)~d.
%Consequently, during the last decades measurements were made repeatedly.
Using Fourier analysis for measurements of circular and linear polarisation, \citet{Bagnulo1995} derived the period of stellar rotation $\emph{P} = 12.461(1)$~d which is in good accordance with the value %derived 
obtained by \citet{Mathys1991}.
%\citet{Preston1972} also noted that the Eu line coincides with the magnetic maximum, while Mg, Ti, Cr, and Fe's lines with the magnetic minimum. 

The extensive spectroscopic and polarimetric study of HD~24712 by \citet{Ryabchikova2007} determined the phase shifts for the different elements according to the pulsation peak between radial velocity and photometric data which allow estimation of the propagation of pulsation maximum. 
The available measurements of the mean longitudinal magnetic field starting from \citet{Preston1972}  up till now are presented in Table~\ref{appendix}. Measurements of the mean longitudinal magnetic field were obtained with the CASPEC Zeeman analyzer by \cite{Mathys1994} which measured the wavelength shifts between right and left circular polarizations. \citet{Leone2000} estimated the mean longitudinal magnetic field by using the same method applied to the polarimetric spectra obtained with the fiber-fed REOSC spectrograph of the Catania Astronomical Observatory. \citet{Wade2000} provided the mean longitudinal magnetic field  derived with high-precision from LSD Stokes V profiles by using the MuSiCoS spectropolarimeter. % least-squares deconvolution to analyze weak lines in spectra. 
\citet{Rusomarov2013} reported with high accuracy 15 new measurements of $<B_{\rm z}>$ (see Table~\ref{appendix}) derived from analysis of LSD Stokes V profiles obtained for iron and rare earth elements by using HARPSpol's polarimetric spectra. From available measurements, these authors determined the parameters for a model of inclined magnetic dipole and determined the rotational period $\emph{P} = 12.45812(19)$~d. %In the proceeding study 
\citet{Rusomarov2015} used magnetic Doppler imaging to provide a detailed magnetic field topology where dipolar components play a key role with a small impact from high-order harmonics.

%Current observations from the \textit{TESS} telescope made it possible to present quite different values. 
Unlike the other stars in our sample, the detrending procedure for HD~24712 was applied separately to each sector. Since the star demonstrates not only rotational modulation but also pulsations inherent to the roAp subclass, we performed prewhitening of high-order frequencies of stellar pulsation as well. From Fourier analysis of \textit{TESS} photometric data combined from Sectors~5 and 31, we detected a signal at the frequency (and its first harmonic) that correspond to the period of rotation $\emph{P} = 12.45862(5)$~d. The period determined in this article appears to be larger by 2~$\sigma$ than to the value obtained by \citet{Rusomarov2015}. The top panel in Fig.~\ref{HD24712} shows a significant signal detected at the low frequencies of the discrete Fourier transform of the reduced light curve. %This frequency is caused by rotational modulation corresponds . 
%The last sentence isn't clear. I don't understand what is meant by "rotational modulation corresponds of the reduced light curve.
%We tried to clarify the period by additional observations of the third cycle of TESS, but due to the large instrumental error, the data were too damaged to use. 
High-frequency roAp oscillations were discovered in HD~24712 by \citet{Kurtz1981}, with the p-mode pulsation period being 6.15~min. %, and significantly improved the accuracy of deviation. 
We have detected p-mode pulsations at high frequencies (see bottom panel in Fig.~\ref{HD24712}) and we determined the pulsation period $\emph{P} = 6.1257(1)$~min  corresponding to the strongest signal. %Hence, as an important part of roAp stars to determine the p-mode pulsations we provide the sufficiently approximate value of the period, P = 6.126$\pm$4$\times$10$^{-5}$~min. 
From the analysis of three frequencies having the most significant amplitudes at the following frequencies 235.07495~d$^{-1}$, 232.19730~d$^{-1}$, 229.20934~d$^{-1}$, we estimated the average separation to be about 2.93281~d$^{-1}$.
%The lower panel in Fig.\ref{HD24712} presents Fourier transform for the high frequencies region with the roAp type. 

\subsection{HD~38170 (TIC~140288359 = WZ~Col)}
\label{HD38170text}

\begin{figure}
\centering
\includegraphics[width=\linewidth]{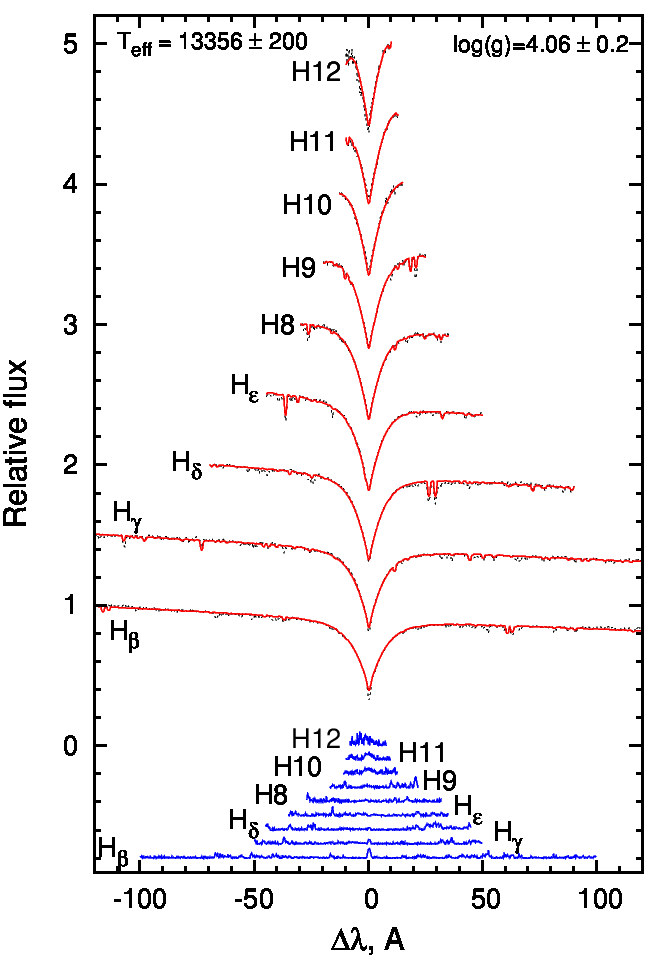}
\caption{The same type of analysis as in Fig.~\ref{figHD22920Bal} except for the synthetic lines of the star HD~63401 were used the value of $T_{\rm eff} = 13400 \pm 200$~K and $\log{g} = 4.1 \pm 0.2$.}
\label{figHD63401Bal}
\end{figure}

\begin{figure}
\centering
\includegraphics[width=\linewidth]{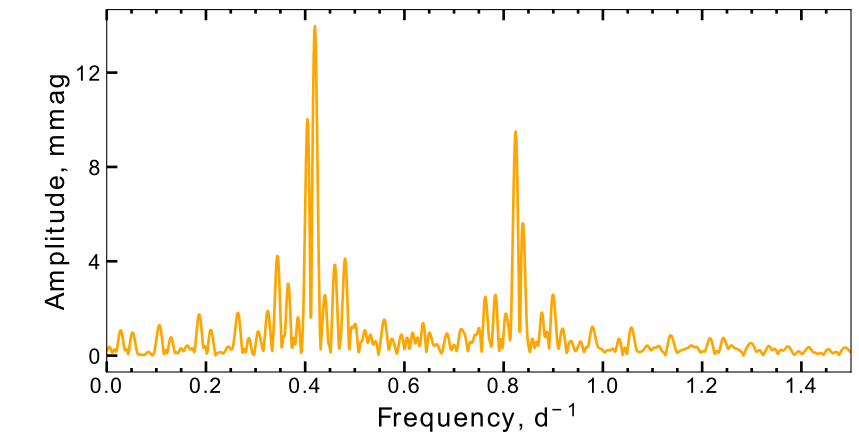}
\includegraphics[width=\linewidth]{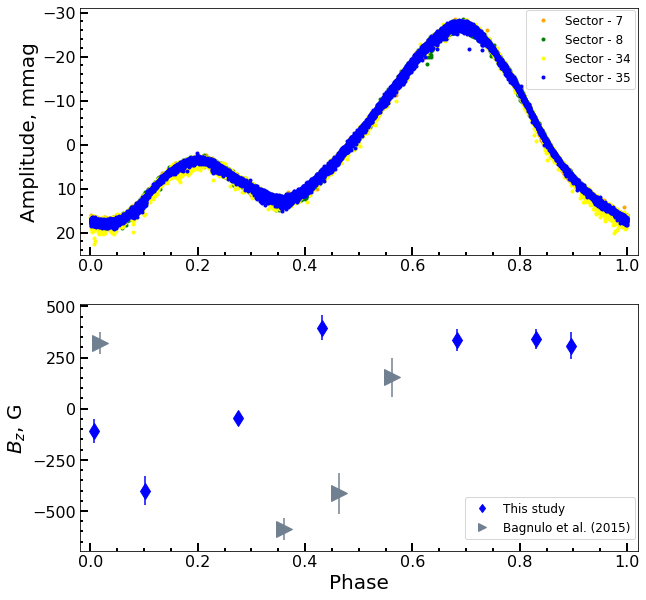}
\caption{ The same as at the Fig~\ref{HD22920rot} but for HD~63401. Phase diagram is built using the derived rotational period $\emph{P} = 2.414474$ for the reduced light curve from the sectors~7,8, and 34,35 and for $<B_{\rm z}>$ measurements from % To plot the image we used the magnetic fields values from observation by 
\citet{Bagnulo2015} (black triangles). }
\label{figHD63401}
\end{figure}

HD~38170 has been identified as a CP star with the spectral class B9 by \citet{Cannon1993}. The radial velocity has been obtained in many studies and the values are clustered around 35~km~s$^{-1}$, but the most recent measurement $36.3 \pm 0.6$~km~s$^{-1}$ has been provided by \citet{Gontcharov2006}.
From the Fourier transform analysis of line profiles, \citet{Royer2002} derived v$\sin{i}$ = $65 \pm 9$~km~s$^{-1}$. %the derived value of v$\sin{i}$=65.0$\pm$9.0~km~s$^{-1}$ derived from the frequency first zero.
%During the study of the evolution of rotational velocities %as a proceeding of previously mentioned investigations, 
\cite{Zorec2012} derived a stellar mass of $M  = 3.07 \pm 0.05 (M_{\sun})$, fractional age on the main sequence $0.89 \pm 0.04$, and other fundamental parameters, such as $T_{\rm eff}$ = $10000 \pm 257$~K and log$(L_{\star}/ L_{\sun}) = 2.17 \pm 0.03$, from % have been determined as fundamental parameters by using 
Str{\"o}mgren-Crawford photometry. Similar mass $M  = 2.8 \pm 0.1 (M_{\sun})$ and age $ 394^{+10}_{-17}$ Myr, which corresponds to log(Age) = $ 8.60^{+0.02}_{-0.04}$ yr, were obtained by \citet{David-Uraz2021} for HD~38170. From the analysis of Balmer line profiles we determined $T_{\rm eff} = 9500 \pm 200$~K and $\log{g} = 3.65 \pm 0.2$ (see Fig.~\ref{HD38170Balmer}).

\citet{David-Uraz2021} confirmed that HD~38170 is a magnetic CP star via direct detection in Stokes V profiles obtained with ESPaDOnS (see Table \ref{appendix}) and calculated its dipolar magnetic strength as $ B_{\rm d} = 254^{+78}_{-49}$~G by applying a Bayesian analysis on LSD profiles \citep{Petit2012}. From the \textit{TESS} first year observations \citet{David-Uraz2021} determined the rotation period $\emph{P} = 2.76618(4)$~d and measured v$\sin{i} = 57 \pm 5$~km~s$^{-1}$. The global stellar parameters obtained in this study for HD~38170 are quite similar to those presented by \citet{David-Uraz2021} considering estimation error bars, except for the mass that is different by more than 1 $\sigma$ (see Table~\ref{Geneva}).

From the \textit{TESS} observations we were able  to identify the rotational period for HD~38170. From Fourier transform analysis (see Fig. \ref{fig38170}) we defined a signal at the low-frequency region that may be interpreted as the first harmonic. We have detrended the light curve of this star obtained by TESS in sectors~5, 6, 32 and 33, and phased it with the period of stellar rotation (see the lower panel of Fig.~\ref{fig38170}). 
On the phase diagram, one can clearly see two maxima that most probably appear due to the presence of abundance spots in the stellar atmosphere of HD~38170. Contrary to the case of  HD~10840 (see Subsection~\ref{HD10840}), the highest peak for HD~38170 is found at larger frequency (second one).% which we considered as the fundamental frequency. 
The rotational period 2.766116(2)~d determined for this frequency is consistent with the result derived by \citet{David-Uraz2021}. 
%Applying the OMR model we suggest that HD~38170 may possess a magnetic field with the best visibility of the magnetic poles.

\subsection{HD~63401 (TIC~175604551 = OX~Pup)}

\begin{figure}
\centering
\includegraphics[width=\linewidth]{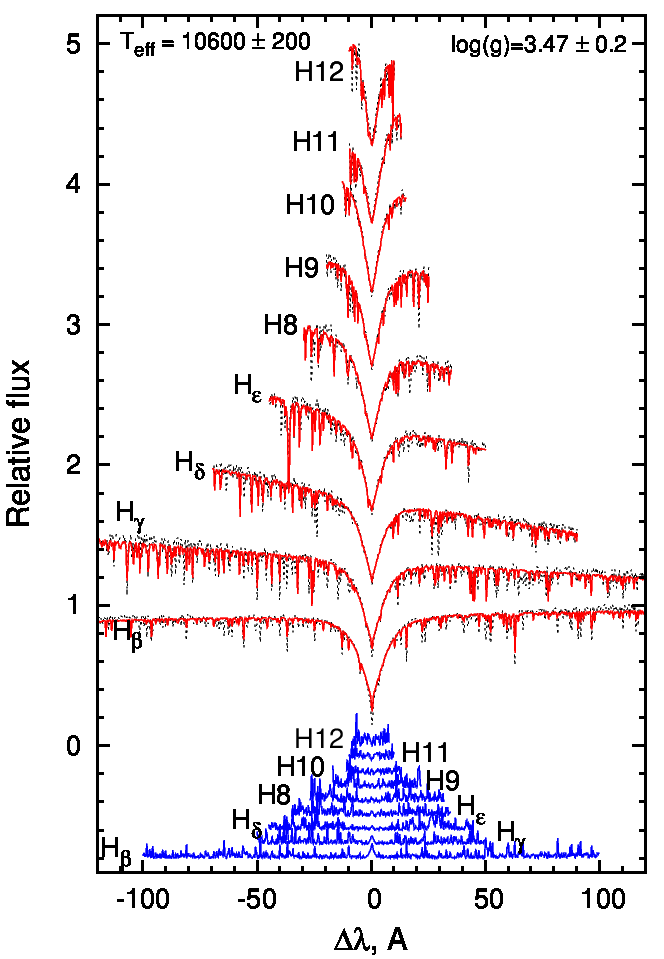}
\caption{The same as in Fig.~\ref{figHD22920Bal} but for best fit of Balmer lines in HD~74521 obtained for  $T_{\rm eff} = 10600 \pm 200$~K and $\log{g} = 3.5 \pm 0.2$.}
\label{figHD74521Bal}
\end{figure}

\begin{figure}
\centering
\includegraphics[width=\linewidth]{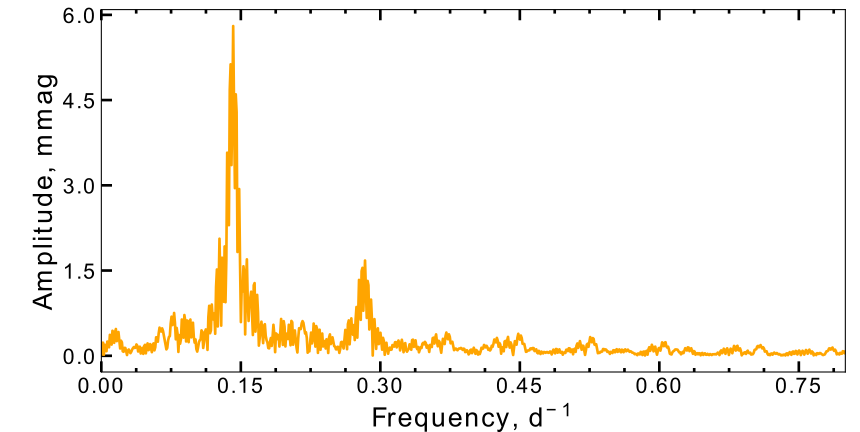}
\includegraphics[width=\linewidth]{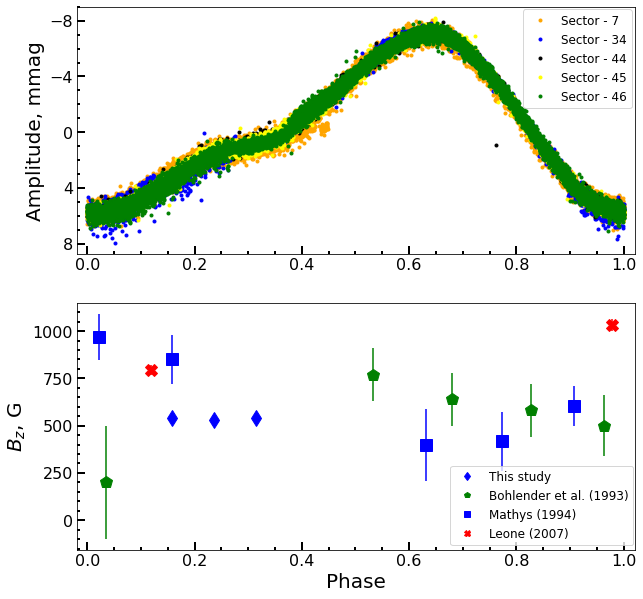}
\caption{The same as at the Fig~\ref{HD22920rot} but for HD~74521. Phase diagram is built using the derived rotational period $\emph{P} = 7.05010$~d for the reduced light curve from the sectors~7, 34 and 44 to 46 and for the $\langle B_{\rm z}\rangle$ measurements from \citet{Bohlender1993} (green pentagons), \citet{Mathys1994} (blue squares and diamonds (private)) and \citet{Leone2007} (red stars).}

\label{figHD74521}
\end{figure}

%In accordance with the extended Henry Draper Catalogue, 
The star HD~63401 was classified as a variable star of spectral type B9 \citep{Cannon1993}.
\citet{Bailey2014} provided for HD~63401, $T_{\rm eff} = 13500 \pm 500$~K which was previously mentioned by \citet{Landstreet2007}, and derived $\log{g} = 4.2 \pm 0.2$ and v$\sin{i} = 52 \pm 4$~km~s$^{-1}$ from the analysis of two FEROS spectra. The radial velocity v$_{\rm r} = 22.0 \pm 1.4$~km~s$^{-1}$ (see Table~\ref{tab1})  was derived for HD~63401 by \citet{Gontcharov2006}. 
%as result of his  global research of v$_{\rm r}$ for stars observed with Hipparcos.
%employing the polarized radiative transfer code \textit{ZEEMAN} \citep{Landstreet1982}. 
We report for HD~63401 similar stellar parameters $T_{\rm eff} = 13400 \pm 200$~K, $\log{g} = 4.1 \pm 0.2$, v$_{\rm r} = 26 \pm 2$~km~s$^{-1}$ and v$\sin{i} = 52 \pm 4$~km~s$^{-1}$ obtained from the fitting of Balmer line profiles in the available ESPaDOnS spectra (see Fig.~\ref{figHD63401Bal}).

%HD~63401 belongs to the set of 15 magnetic Bp stars found in the open clusters for which \citet{Bailey2014} detected an overabundance of He, O, Mg, Si, Ti, Cr, Fe, Pr, and Nd. %, some of these elements have a strong overabundance in the atmosphere. 
 % stars observation \cite{Gontcharov2006}.

\citet{Glagolevskij2007} studied HD~63401
%in the framework of the project ''Magnetic-field dependence of Chemical Anomalies in CP Stars''
and determined the mean longitudinal magnetic field (line-of-sight component) $<B_{\rm e}> = 70\pm50$~G and the root-mean-square magnetic field $B_{\rm rms} = 400$~G. %Following, 
\citet{Landstreet2007} derived the value $B_{\rm rms} = 365$~G, which is in good accordance with the result obtained by \citet{Glagolevskij2007}. \citet{Bagnulo2015} used polarimetric spectra obtained with the spectrograph FORS1, and measured the mean longitudinal magnetic field from the analysis of circular polarisation in the Balmer line profiles and metal lines (see first column in Table~\ref{appendix}). 

The period of stellar rotation $2.41 \pm 0.02$~d was derived from the analysis of the light curves observed at ESO, with the ESO 50 cm telescope and the Danish 50 cm telescope, and reported for the first time by \citet{Hensberge1976}. From the available photometric TESS data, we confirm the value of the above period. The top panel of the Fig.~\ref{figHD63401} presents the discrete Fourier transform of the reduced light curve observed by \textit{TESS} in sectors~7, 8, 34 and 35. The significant signals detected at low frequency and its first harmonic may be caused by stellar rotation with the period $\emph{P} = 2.414474(1)$~d. In the middle panel in Fig.~\ref{figHD63401}, we show the phased light curve with period of rotation determined from reduced \textit{TESS} data. Looking for a precise period of rotation, we applied Fourier transform to combined $<B_{\rm z}>$ measurements from published data \citep{Bagnulo2015} and this study in the vicinity of the fundamental rotational frequency determined from the \textit{TESS} data. This attempt to obtain a more precise period was motivated by the observed phase shift between two $<B_{\rm z}>$ datasets when folded with the period derived from \textit{TESS} data (see bottom panel in Fig.~\ref{figHD63401}). In result, the proper phasing of magnetic field measurements using the photometric period was unsuccessful.% and, in the bottom panel the phase diagram for the available $<B_{\rm z}>$ data.}% to which was applied rotational period $\emph{P} = 2.41414(6)$~d extracted from the $<B_{\rm z}>$ measurements.} %Assuming the OMR model for HD~63401 we suggest that light curve modulation is caused by different visibility of overabundance spots located close to the magnetic poles. %if both graphs have a maximum and minimum position at the same phase the star has overabundance at the poles in the atmosphere produced by the magnetic field.

\subsection{HD~74521 (TIC~443995718 = BI~Cnc)}

\begin{figure}
\centering
\includegraphics[width=\linewidth]{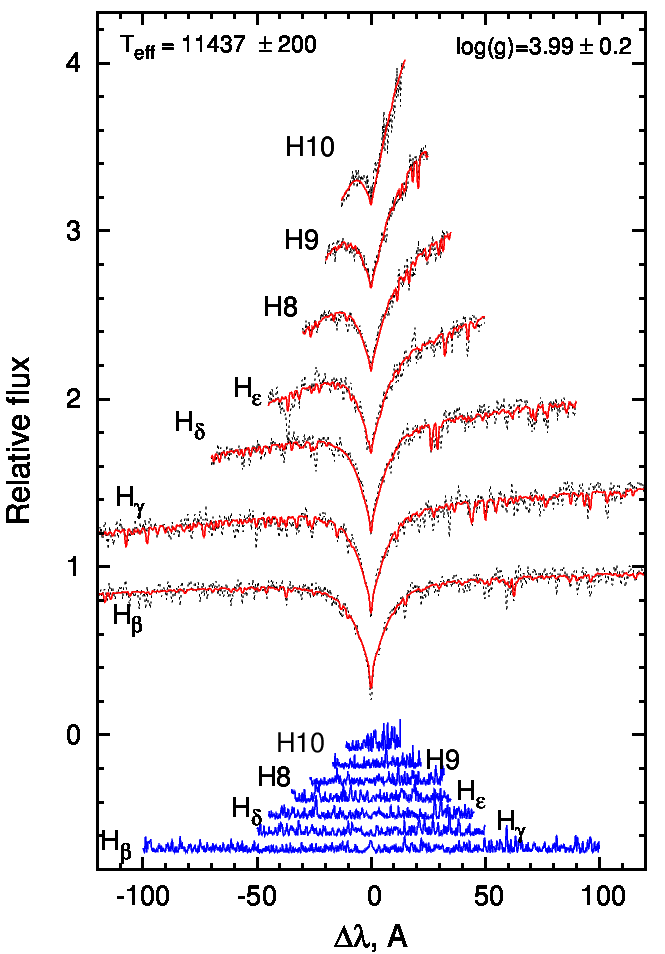}
\caption{The same type of analysis as in Fig.~\ref{figHD22920Bal} except for the synthetic lines of the star HD~77314 were used the value of $T_{\rm eff} = 11400 \pm 200$~K and $\log{g} = 4.0 \pm 0.2$. }
\label{HD77314Balmer}
\end{figure}

\begin{figure}
\centering
\includegraphics[width=\linewidth]{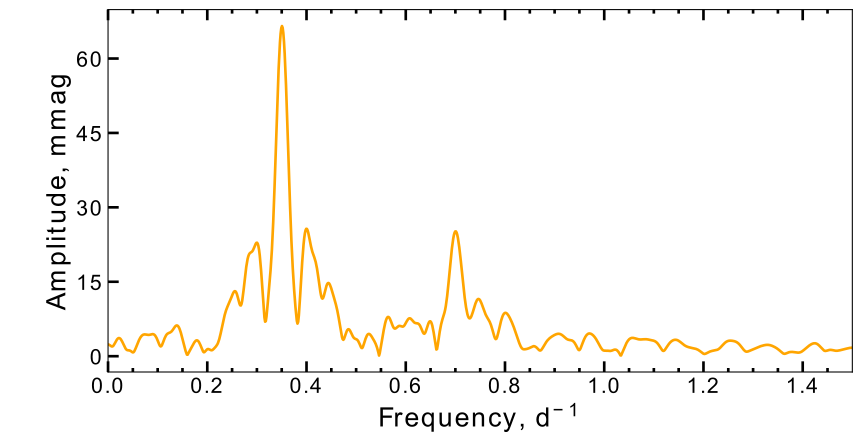}
\includegraphics[width=\linewidth]{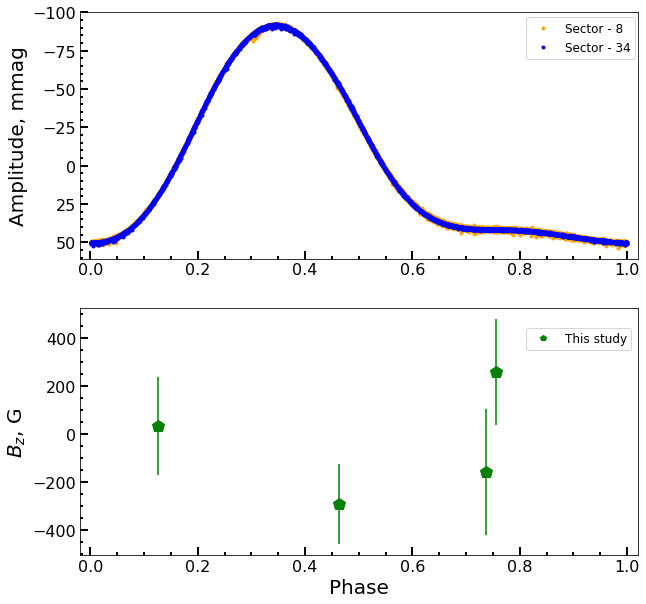}
\caption{The same as at the Fig~\ref{HD22920rot} but for HD~77314. The phase diagram is built using the derived rotational period $\emph{P} = 2.864325$ for the reduced light curve from the sectors~8 and 34 and for $<B_{\rm z}>$ DAO longitudinal magnetic field measurements (green pentagons).}
\label{figHD77314}
\end{figure}

%In the process of researching of chemically peculiar Ap and Am stars, \citet{Abt1995} noted that stars with rapid rotation have normal spectra compare to slowly rotating stars with abnormal spectra. 
\citet{Abt1995} provided the spectral type A1Vp for HD~74521 with HgMnSiEu strong lines and CaMg weak lines, and v$\sin{i} = 10$~km~s$^{-1}$. \citet{Royer2002} used Fourier transform analysis of line profiles in the range 420-460~nm to combine the set of rotational velocities for Ap stars on the basis of the previous study of \citet{Abt1995}, and derived for HD~74521  v$\sin{i} = 18.0 \pm 0.9$~km~s$^{-1}$, and \citet{Mathys1995} determined v$\sin{i} = 19.0 \pm 4.6$~km~s$^{-1}$. \citet{Gontcharov2006} provided the value of radial velocity v$_{\rm r} = 27.5 \pm 1.4$~km~s$^{-1}$. \citet{Wraight2012} provided for HD~74521 an effective temperature of $T_{\rm eff} = 10789 \pm 500$~K. Our estimates of $T_{\rm eff} = 10600 \pm 200$~K derived from the fitting of Balmer lines (see Fig.~\ref{figHD74521Bal}) and $T_{\rm eff} = 10500 \pm 300$~K inferred from the Str{\"o}mgren-Crawford photometry (see Table~\ref{Geneva}) are consistent with the results of \citet{Wraight2012} considering the estimated uncertainties. Also, from the analysis of Balmer line profiles we have obtained v$\sin{i} = 18 \pm 3$~km~s$^{-1}$ and v$_{\rm r} = 24 \pm 3$~km~s$^{-1}$ which are in good accordance with the previously published data taking into account the estimation error bars (see Table~\ref{tab1}).

%\citet{Wraight2012} analyzed light curves of the chemically peculiar stars observed with STEREO spacecraft to study the flux variability influenced by a significant magnetic field. 

HD~74521 was investigated as a mCP star since \citet{Stepien1968} who measured the period of stellar rotation to be $\emph{P} = 5.43$~d. During the last decades, many rotation periods were derived for HD~74521 from terrestrial and space observations. Recent measurements have been obtained by \citet{Leone2007} as $\emph{P} = 7.0486(5)$~d, $\emph{P} = 6.91(1)$~d by \citet{Wraight2012} and  $\emph{P} =7.0501(2)$~d by \citet{Dukes2018}. 

\textit{TESS} has carried out observations of HD~74521 in sectors~7, 34 and 44 to 46,  which provide a valuable time baseline of almost 3~yr. From the discrete Fourier transform of the reduced light curve, we confirm the value of the rotational period to be $\emph{P} = 7.05010(5)$~d. The top panel of Fig.~\ref{figHD74521} shows a significant signal at $\nu = 0.1418419(1)$~d$^{-1}$ that is related to stellar rotation. 
%aking into account the results presented (see top panel of Fig\ref{figHD74521}) we identified a signal (first harmonic is $\nu$ = 0.1418419(1)
%(2) its uncertainty 
%~d$^{-1}$) in the region of low frequencies. 
The middle and bottom panels of Fig.~\ref{figHD74521} present a phase diagram. The measurements of the mean longitudinal magnetic field are taken from  \citet{Bohlender1993}, \citet{Mathys1994} and \citet{Leone2007}. The mean longitudinal magnetic field was measured with the Pockels cell polarimeter and the CASPEC Zeeman analyzer respectively. \citet{Leone2007} measured the mean longitudinal magnetic field across the whole spectrum and through Balmer lines obtained with the spectropolarimeter ISIS. New three $<B_{\rm z}>$ measurements were derived for this object using high-resolution spectropolarimetric data recently obtained with ESPaDOnS. Even though the mean longitudinal magnetic field is consistently detected at the level of several hundred G, different $<B_{\rm z}>$ datasets show systematic differences with very low or approaching zero variability in each.

%Considering the OMR model we assume that the abundance anomalies in the atmosphere of HD74521 are partially caused by the contribution of the relatively strong magnetic field. 
%The $\langle B_{\rm z}\rangle$ measurements %seem to be correlated with the light curve phase diagram 
%variate on the phase diagram with $\emph{P} = 7.05010(5)$~d and are approaching its extremum at the minimum of the phased light curve (see bottom panel of Fig.~\ref{figHD74521}). 
%tare correlated with the phase of stellar rotation furthermore the poles concur with the maximum and minimum phase.
\begin{figure}
\centering
\includegraphics[width=\linewidth]{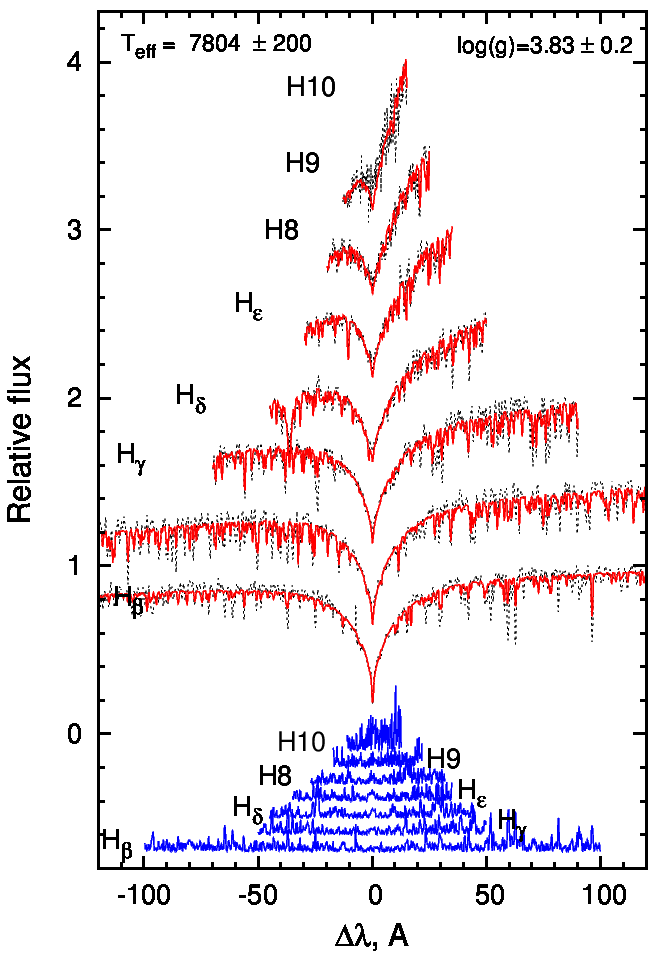}
\caption{The same as in Fig.~\ref{figHD22920Bal} but for best fit of Balmer lines in HD~86592 obtained for $T_{\rm eff} = 7800 \pm 200$~K and $\log{g} = 3.8 \pm 0.2$.}
\label{figHD86592Bal}
\end{figure}

\subsection{HD~77314 (TIC~270487298 = NP~Hya)}

%Based on the data from the Henry Draper catalog of \citet{Cannon1993}, 
HD~77314 is a CP star of spectral type A2 \citep{Cannon1993}, or alternative type ApSrCrEu \citep{Houk1999}.
%is assigned to it in the Michigan catalog of \citet{Houk1999}. 
The \textit{TESS} input catalogue (TIC) provides the effective temperature $T_{\rm eff}= 9253 \pm 372$~K and surface gravity $\log{g} =3.69 \pm 0.10$ \citep{Stassun2018, Stassun2019}. From the fitting of Balmer line profiles in the HERMES spectra (see Fig \ref{HD77314Balmer}) we have derived $T_{\rm eff} = 11400 \pm 200$~K and $\log{g} = 4.0 \pm 0.2$ %that considerably diverge from the ones presented in the 
(see Tables~\ref{tab1},~\ref{Geneva}). As there was no data available in the Geneva and Str{\"o}mgren-Crawford photometry catalogs, we could not confirm any of the above data, which are in stark disagreement.

From the extensive study of rotational periods from MASCARA data, \citet{Bernhard2020} derived the rotation period of $\emph{P} = 2.86445(8)$~d which is in a good accordance with the period of $\emph{P} = 2.8646$~d presented in the International Variable Star Index of the American Association of Variable Star Observers \citep[VSX;][]{Watson2006}.
From the discrete Fourier transform of the detrended light curve observed in sectors~8 (2~min cadence) and 34 (10~min cadence) (see top panel Fig~\ref{figHD77314}) we have detected the signals in the low-frequency region that most probably correspond to the rotation period %. Based on the above we provide the period of stellar rotation
$\emph{P} = 2.864325(1)$~d which %in good accordance with
 deviates from the period determined by \citet{Bernhard2020} by more than 1~$\sigma$. In the bottom panels of Fig~\ref{figHD77314}, we provide the phase diagram built with this period for the light curve and  for the measurements of the mean longitudinal magnetic field. Considering the obtained error bars the derived $\langle B_{\rm z}\rangle$ measurements do not appear to differ significantly from zero. We need to use Stokes I and V spectra with higher signal to noise ratio for HD~77314 (e.g. from ESPaDOnS) to measure its magnetic field with high precision.
%provided by D.Bohlender. The measurements of the magnetic field were implemented and calculated $\langle B_{\rm z}\rangle$ by D.Bohlender (see Tab~\ref{HD77314_Bz}). 

%Taking into account the OMR model, we suggest that the pole of the magnetic field corresponds to the peak in the phase diagram. 

\subsection{HD~86592 (TIC~332654682 = V359~Hya)}

\citet{Houk1988} determined the spectral type of HD~86592 as ApSrEuCr. In this study, the effective temperature $T_{\rm eff} = 7800 \pm 200$~K and surface gravity $\log{g} = 3.8 \pm 0.2$ are estimated for HD~86592 from the fitting of Balmer line profiles observed with HERMES (see Fig.~\ref{figHD86592Bal}). Considering the estimated uncertainties these data are consistent with the values of $T_{\rm eff} = 7955 \pm 95$~K and $\log{g} = 4.28 \pm 0.18$ from \citet{Kordopatis2013} and to our estimates of $T_{\rm eff} = 7700 \pm 300$~K and $\log{g} = 4.08 \pm 0.06$ obtained for this star from the Geneva photometry (see Table~\ref{Geneva}). 

The presence of a strong magnetic field did not allow \citet{Babel1997} to estimate the value of rotational velocity  from individual line profiles, therefore, they used the correlation dip and derived the values v$_{\rm r} = 12.7 \pm 0.3$~km~s$^{-1}$ and v$\sin{i} = 16 \pm 2$~km~s$^{-1}$ for HD~86592. Our estimate of the radial velocity v$_{\rm r} = 13 \pm 1$~km~s$^{-1}$ is in a good accordance with the published value, while v$\sin{i} = 27 \pm 5$~km~s$^{-1}$ appears to be overestimated due to the magnetic widening of line profiles and other line broadening factors. 
%The period of the stellar rotation 2.8867~d. had been defined from photometric measurements in the Geneva system. 
The study of magnetic fields of cool CP stars leads \citet{Babel1997} to discover %the strong magnetic field in HD~86592. %Measurements of the mean longitudinal magnetic field have been carried out using the  cross-correlation method applied to the analysis of ELODIE spectra. %and in principle used the cross-correlation method. Thus \citet{Babel1997} had produced 
%These authors also measured a value of
the square of the quadratic magnetic field $\sqrt{\langle B^2\rangle}$\footnote{Note the replacement of $H$ by $B$, for consistency with the field notation used in this paper} = 15.5-16.1~kG by the method described in \citet{Mathys1995} applied to ELODIE spectr. 
%the quadratic magnetic field $\langle H^{2} \rangle = 15.5-16.1$~kG by the method described in \citet{Mathys1995} applied to ELODIE spectra. 

\begin{figure}
\centering
\includegraphics[width=\linewidth]{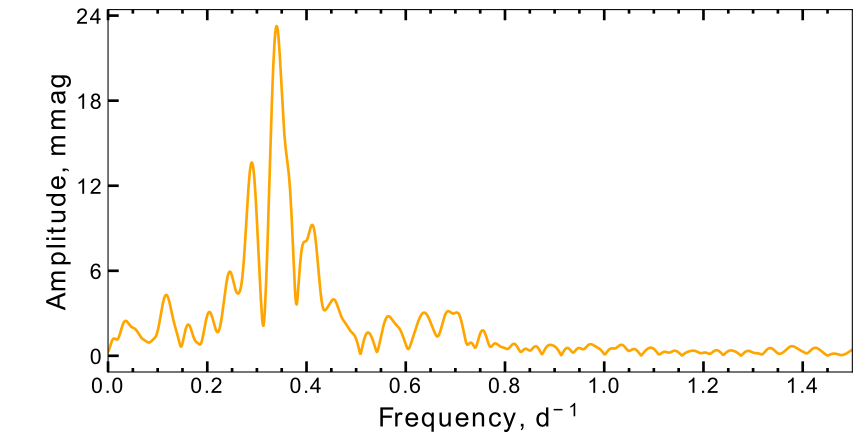}
\includegraphics[width=\linewidth]{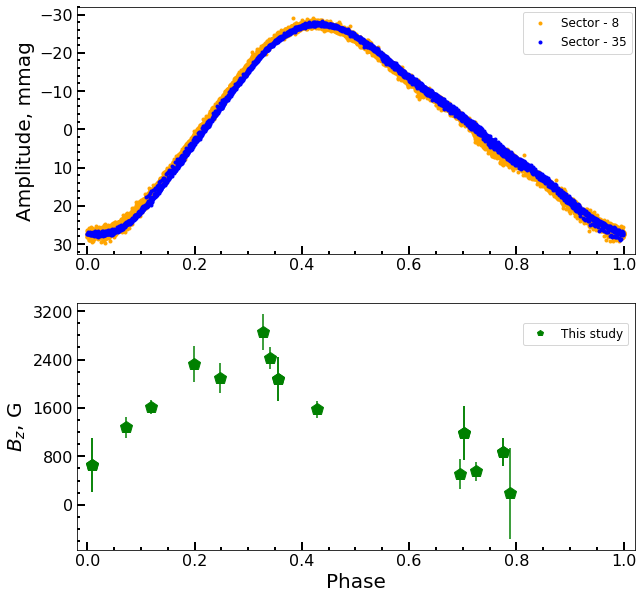}
\caption{The same as at the Fig.~\ref{HD22920rot} but for HD~86592. Phase diagram is built using the derived rotational period $\emph{P} = 2.88657$ for the reduced light curve from the sectors~8 and 35 and for $<B_{\rm z}>$ DAO  longitudinal magnetic field measurements(green pentagons).}
\label{figHD86592}
\end{figure}

%Otherwise than the rest of the stars, for this one we used LC extracted from \textit{TESS} Science Processing Operations Center (SPOC) pipeline \citep{Jenkins2016}. This decision was made based on the previously analyzed from MAST and MIT’s Quick-Look Pipeline (QLP) data. Since MAST does not provide data from Sector 35 and it was a major source, mixing data lead to a shifted period of rotation with precision second digit after the comma. Different pipelines introduce addition noise since the same observed image leads to different flux (LC), especially visible for this star in Sector 8.  
From the analysis of the discrete Fourier transform of the detrended light curve observed by \textit{TESS} in sectors~8 and 35 we have derived for HD~86592 a rotation period of $\emph{P} = 2.88657(3)$~d (top panel in Fig.~\ref{figHD86592}). 
%The detected signal (see top panel at Fig.~\ref{figHD86592}) was identified as the first harmonic that correlated to the stellar rotation of the star. 
The obtained period coincides with the period P= 2.8867~d provided by \citet{Babel1997} from photometric measurements carried out in the Geneva system. In the middle and bottom panels of Fig.~\ref{figHD86592}, we provide the light curve and the measurements of the mean longitudinal magnetic field phased with the period of stellar rotation $\emph{P} = 2.88657(3)$~d derived for in this study. The available measurements of $\langle B_{\rm z}\rangle$ have been obtained with the help of the dimaPol spectropolarimeter %\citep{Monin2015} 
and resulted in the period $\emph{P} =2.887318(2)$~d
%(Bohlender, private communication)
which is very close to the one derived from the \textit{TESS} photometry.  %For magnetic field we used the private data which was measured through a spectropolarimeter by D.Bohlender and provide period of stellar rotation 2.887424~d. 
In the bottom panel of Fig.~\ref{figHD86592}, one can see that the mean longitudinal magnetic field reaches a probable extremum at the phase $\varphi = 0.3$ which is quite close to the maximum of the reduced light curve ($\varphi= 0.4$). 
%In the frame of the OMR model, magnetic field contributes to the horizontal stratification of elements' abundance that in turn causes rotational modulation of the observed light curve for HD~86592. %for the description of the behavior light variability as vertical and horizontal inhomogeneous in the presence of the magnetic field.

\section{Discussion}
\label{discus}

\begin{figure}
\centering
\includegraphics[width=\linewidth]{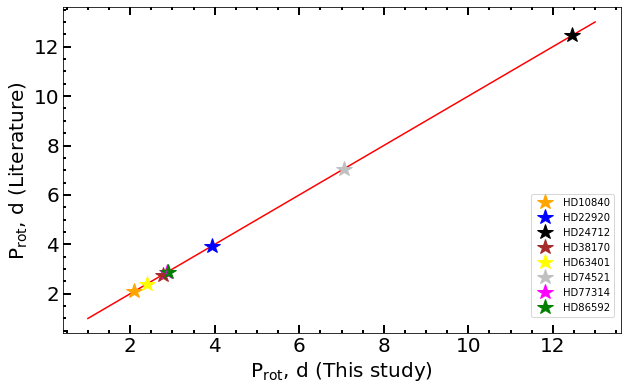}
\caption{P$_{\textrm {rot}}$ presented in this study vs P$_{\textrm {rot}}$ extracted from literature (see Table~\ref{tab1}). Red solid line of $x = y$ represents consistency of derived results. Uncertainties are smaller than the used symbols considering the high precision of period estimates.}
\label{ProtvsProt}
\end{figure}

We carried out analysis of eight Ap stars with the purpose to determine rotation periods from the recently conducted observations with the NASA \textit{TESS} space telescope and confirm the existence of rotational modulation by comparing phased light curve with the magnetic field measurements phased, with the same period. Thus we have collected all available data for each star from the MAST database up to 50 sectors.  
%check up to 50 sectors available
%at this moment. 
We have also collected all available measurements of the mean longitudinal magnetic field $\langle B_{\rm z}\rangle$ for each star (see Table~\ref{appendix}), and phased them according to the derived period. % derived in this study. 
Based on the obtained data and the detrended light curves
%, light curves after the detrending procedure, we can state 
one can admit that the periods derived in this study coincide with those previously mentioned in the literature (see Fig.~\ref{ProtvsProt}). Nevertheless in most cases, our results are obtained with a much higher precision provided by much larger timescale of %recording
analyzed data. 

\begin{figure}
\centering
\includegraphics[width=\linewidth]{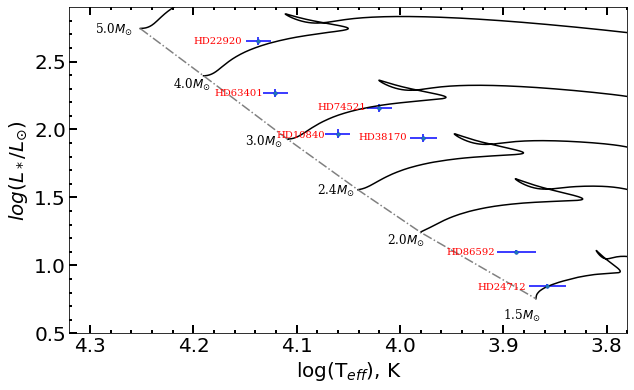}
\includegraphics[width=\linewidth]{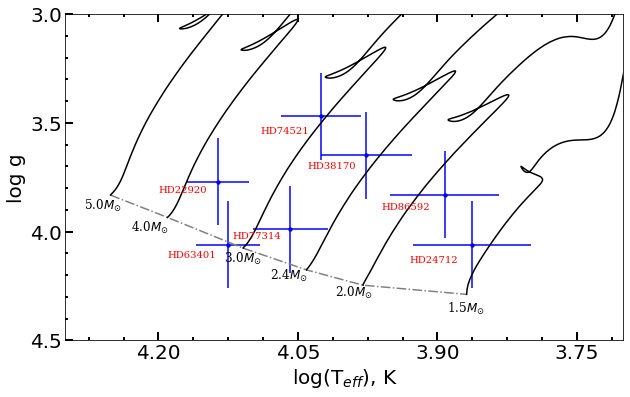}
\caption{Hertzsprung–Russell diagram (top panel) and Kiel diagram (bottom panel) built for 7 stars from the studied sample. Evolutionary tracks are calculated for stars of 1.5-5.0 $M_{\sun}$ assuming solar metallicity without rotation (the continuous lines) starting from the zero age main sequence (dotted line) \citep{Bressan2012}}
\label{fig_HR_diagram}
\end{figure}

To illustrate the diversity of the global stellar parameters we built Hertzsprung Russell (see top panel of Fig. \ref{fig_HR_diagram}) and Kiel (see bottom panel of Fig. \ref{fig_HR_diagram}) diagrams for 7 targets from our sample by using their derived values of effective temperature and luminosity from Geneva and Str{\"o}mgren-Crawford photometry and $\log{g}$ from the analysis of Balmer line profiles. %(see Table~\ref{Geneva}), and by overlapping with 
We used evolutionary curves of solar metallicity $Z = 0.014$ without rotation for stars with 1.5-5.0 $M_{\sun}$ \citep{Bressan2012}, also $\log{g}$ for evolutionary curves were calculated by using formulae derived by \citet{Netopil2017}. HD~77314 was not included in the HR diagram and HD~10840 was not included in the Kiel diagram because of absence of observational data required to derive stellar parameters used to build these diagrams. Nevertheless, position of studied stars on these diagrams is consistent with evolutionary tracks of \citep{Bressan2012} calculated for the same stellar mass, with exception of HD~22920 shifted towards a lower mass level in the Kiel diagram (see Fig~\ref{fig_HR_diagram}).

\begin{figure}
\centering
\includegraphics[width=\linewidth]{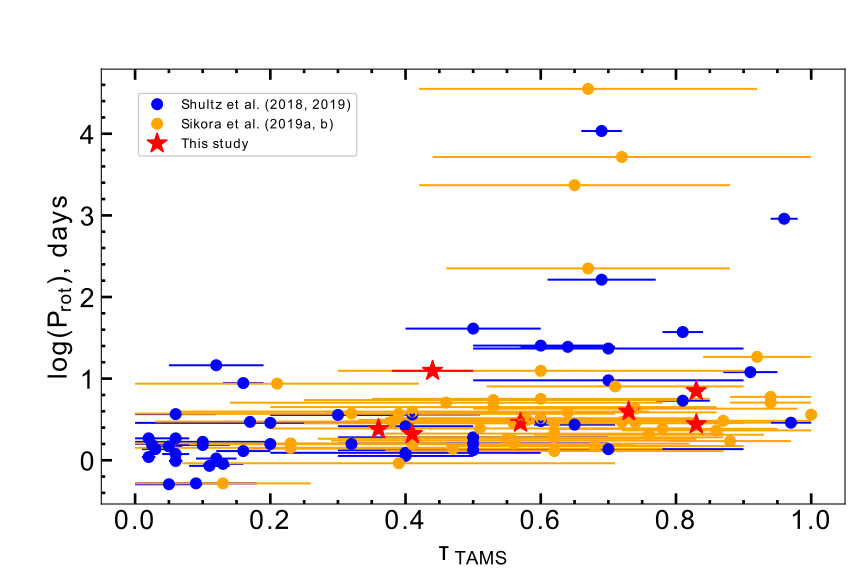}
\includegraphics[width=\linewidth]{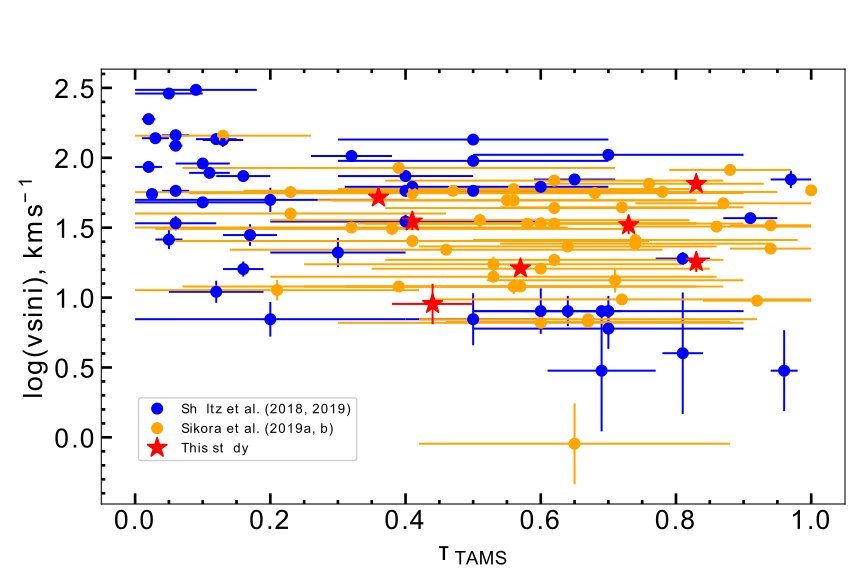}
\caption{Rotational period (upper panel) and v$\sin{i}$ (lower panel) as a function of fractional age for the sample selected for this study (red stars, see Tab.~\ref{Geneva}) and for the samples of ApBp stars analysed by \citet{Sikora2019_2, Sikora2019b} (orange filled circles) and  \citet{Shultz2018, Shultz2019} (blue filled circles).
}
\label{figDisscussion}
\end{figure}

Strong dipolar magnetic fields of fossil origin influence rotational rate in massive, hot OB type stars and 
%describe in term of magnetic braking aka 
causes a loss of angular momentum. \citet{Keszthelyi2020} studied evolutionary models of magnetic stars with dipolar magnetic fields, implementing a magnetic braking formula calibrated from the %taking into account 
magnetohydrodynamic simulation \citep{Ud-Doula2009}. %applied for dipolar magnetic fields. 
The derived evolutionary model that accounts for magnetic braking was tested on the experimental measurements \citep{Shultz2018} and exhibits rotational spin down of Bp stars. Therefore, we aspire to test whether the ages and periods determined in this study are consistent with those derived for a much larger sample. %population. 
We used an extensive sample of 100 objects to look for a relation between the fractional main-sequence age ${\tau}_\mathrm{TAMS}$ and the rotational rate $v_{\rm eq}/v_{\rm crit}$ %build this image 
by employing data from the studies of rotational properties of ApBp stars by \citet{Sikora2019_2, Sikora2019b} (45 objects) and by \citet{Shultz2018, Shultz2019} (48 objects), plus seven stars from this study for which we derived their age (see Tab.~\ref{Geneva}).
Fractional main-sequence ages ${\tau}_\mathrm{TAMS}$ for the objects from this study were derived from evolutionary tracks provided by \citet{Bressan2012}\footnote{https://people.sissa.it/~sbressan/parsec.html}. Fig.~\ref{figDisscussion} presents a dependence of the rotational period (top panel) and v$\sin{i}$ (bottom panel) on the fractional main-sequence age. While the derived periods of stellar rotation seem to increase with the fractional age (see top panel of Fig.~\ref{figDisscussion}), the measured values of v$\sin{i}$ show a tendency to decrease with ${\tau}_\mathrm{TAMS}$ %the stellar fractional age 
(see bottom panel of Fig.~\ref{figDisscussion}). 
These data argue in favour of loss of the angular momentum by the magnetic ApBp stars with their age even if some targets from our sample possess different rotational periods at the similar ${\tau}_\mathrm{TAMS}$. 

A dependence of the critical rotational fraction $v_{\rm eq}/v_{\rm crit}$ %and the period of stellar rotation
on the fractional age of studied object at the main sequence ${\tau}_\mathrm{TAMS}$ is presented in Fig.~\ref{fig_Vcrit_Age}. Based on the available stellar parameters we determined $v_{\rm eq}/v_{\rm crit}$ for every star in the sample (see Equation~\ref{veqvcrit}). 
%the bottom panel of Fig.~\ref{figDisscussion}. The top panel of 
This figure shows clearly a decrease of the critical rotational fraction $v_{\rm eq}/v_{\rm crit}$ with the fractional age. The obtained value of the slope stands at -12.0$\pm$ 3.4 which appears to be statistically significant at the 3~$\sigma$-level. %From the image presented on Fig.\ref{figDisscussion} (top panel) we can see the pattern between age and the $v_{\rm eq}/v_{\rm crit}$, which we decided to describe by a linear approximation.
We have presented the prediction interval that describes deviation of our estimates of rotational rates from their least square linear approximation. The data may overcome the confidence interval \citep{Cosma2019}, nevertheless, in our case, only HD~38170 and HD~24712 deviate further from the confidence interval (see pink area in Fig.~\ref{fig_Vcrit_Age}) 
which proves that the distribution of rotation in our sample is consistent with that observed in the broader population of ApBp stars. %Extensive sample used for analysis provide the result which is statistically significant with values $p < 0.05$ and ${R}^2$ = 0.11.  
%The correlation between the 
The derived decrease of the ratio $v_{\rm eq}/v_{\rm crit}$ %and 
with the fractional age seems to be statistically significant (though does not explain much of variation of the data) with the values $p < 0.05$ and ${R}^2$ = 0.12, and means that younger magnetic ApBp stars rotate faster than older ones on the main sequence.

Our finding is in a good agreement with the works of \citet{Abt1979} and \citet{Wolff1981} arguing that the magnetic ApBp stars start losing their angular momentum after they reach the main sequence. Recent studies of \citet{Shultz2019} have confirmed a slow down of stellar rotation of early Bp stars with age and demonstrated that a sample of Ap stars \citet{Sikora2019_2, Sikora2019b} behave in a similar fashion, which is associated with magnetospheric braking.

A similar increase of rotational period for ApBp stars with their MS age has been found by \citet{Kochukhov2006} from the analysis of a sample of 194 objects. This trend appears to be significant for the massive CP stars ($M_{\star} > 2 M_{\sun}$) and for the low mass objects ($M_{\star} \leq 2 M_{\sun}$) which is consistent with our results (see top panel in Fig.~\ref{figDisscussion}). \citet{Kochukhov2006} have reported that for the ApBp stars with $M_{\star} > 3 M_{\sun}$  the observed increase of rotational period can be explained by the changes of the moment of inertia, while for the group ApBp stars with masses withing  $2 M_{\sun} < M_{\star} \leq 3 M_{\sun}$ this trend may be partially caused by the loss of the angular momentum. Similar results for the massive CP stars ($M_{\star} > 3 M_{\sun}$) have been also obtained by \citet{Hubrig2007}.
%On other hand,
During investigation of an extensive sample of CP2 and CP4 stars, \citet{Netopil2017} deduced that the angular momentum of CP stars is conserved throughout their main sequence life. %, without any evidence of additional magnetic braking during this time.
%If one consider a much smaller sample of only CP2 stars (the magnetic Bp/Ap stars, see Fig.~9 in \citet{Netopil2017}).  
%Note that the 
These authors noted however that the theoretical 
%period change related to the conservation of angular momentum on the MS is moderate 
change of period with surface gravity $\log{g}$, which is used as the evolutionary parameter \citep{North1998}, is moderate %(a factor of 2 or less), 
and cannot account for the %huge spread in the 
significant distribution of rotational periods observed in ApBp stars,
%of the Ap star rotation periods, 
especially for the most slowly rotating ones which is consistent with the results of \citet{Kochukhov2006}.
%super-slow rotation of a considerable fraction of them. The latter can only be accounted for as the result of pre-main-sequence evolution. 
\citet{Netopil2017} emphasized that the very-slowly rotating ApBp stars and most of the slowly rotating ApBp stars possess a relatively strong magnetic field implying the crucial role of magnetic field in the slowing-down the rotation of these objects.
%anticorrelates with the rotation velocity of the star as a star with a stronger magnetic field rotates slower.}%As a result, this process leads to weakening of the magnetic field in Bp stars and to the simplification of magnetic field geometry. 

\section{Summary}
\label{summary}

\begin{figure}
\centering
\includegraphics[width=\linewidth]{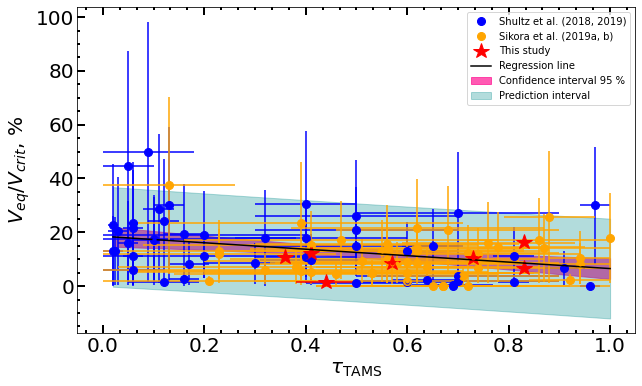}
\caption{Rotational rates $v_{\rm eq}/v_{\rm crit}$ as a function of fractional age for the seven studied stars (red stars, see  Tab.\ref{Geneva}) and for the ApBp stars reported in  \citet{Sikora2019_2, Sikora2019b} (orange filled circles) and  \citet{Shultz2018, Shultz2019} (blue filled circles). 
}
\label{fig_Vcrit_Age}
\end{figure}

In this paper, we aimed to test the detrending methods applied to the light curves with rotational modulation provided by \textit{TESS} for a sample of magnetic ApBp stars with a fairly wide range of global stellar parameters (see Tables~\ref{tab1} \& \ref{Geneva}).
%of data processing observed with \textit{TESS} for selected targets in a fairly wide range of sampling criteria. 
One of our selected targets, HD~24712, does show high-overtone stellar pulsations of roAp type. 
%Among the selected objects, we had the vast majority of Ap stars that show rotational modulation, as well as %a roAp star
%stellar pulsations of roAp type - HD~24712. 
Discrete Fourier transform of the detrended light curves and available measurements of the mean longitudinal magnetic field (see Table~\ref{appendix}) allowed us to derive with high-precision periods of stellar rotation for the selected targets, which are in a good accordance with the previously published data (see. Section~\ref{results}). Analysis of the stellar global parameters derived from the fitting of Balmer line profiles (see Tables~\ref{tab1}) and from the Str{\"o}mgren-Crawford photometry (see Tables~\ref{Geneva}) in combination with the analysis of data from \citet{Sikora2019_2, Sikora2019b} and \citet{Shultz2018, Shultz2019}
resulted in detection of clear decrease of the rotational rate $v_{\rm eq}/v_{\rm crit}$ with the fractional main-sequence age ${\tau}_\mathrm{TAMS}$ %stellar age
(see Fig.~\ref{fig_Vcrit_Age}) indicating the loss of angular momentum of ApBp stars with their evolution on the main sequence \citep{Abt1979, Wolff1981}. %{\bf This result seems to be in contradiction to the recent finding of \citet{Hubrig2007}.  Meanwhile, 
Considering that our sample contains a considerable number of intermediate mass ApBp stars the found indications of the angular momentum loss are consistent with the arguments for a such possibility expressed by \citet{Kochukhov2006} and \citet{Netopil2017}.
The sample analysed in this study includes 100 ApBp stars and is smaller comparing to the samples used by the aforementioned two groups \citep{Kochukhov2006, Netopil2017}.
%To verify our result we plan to engage MOBSTER collaboration\footnote{https://mobster-collab.com/} to improve statistics and 
As a part of MOBSTER collaboration\footnote{https://mobster-collab.com/} we aim to verify our result by expanding 
%For the further plan of investigation 
%we plan 
%to } expand
significantly the sample of magnetic CP stars with rotational modulation towards low and intermediate mass ($M_{\star} \leq 3 M_{\sun}$) objects by using the photometric data obtained from \textit{TESS}  and hopefully from the space telescope \textit{PLATO} as well.
%the studied objects of class Ap with the purpose of studying of dependence of change of the period with age.

\section*{Acknowledgments}
We are grateful to Dmitry Monin for processing data from {\em dimaPol} observations at DAO.
O.K. and V.K. are thankful to the Facult\'{e} des \'{E}tudes Sup\'{e}rieures et
de la Recherch and to the Facult\'{e} des Sciences de l'Universit\'{e} de
Moncton for the financial support of this research. 
V.K. and C.L.
%G.A.W.
acknowledge support from the Natural Sciences and Engineering Research Council
of Canada (NSERC). A.D.-U. is supported by NASA under award number 80GSFC21M0002. D.M.B. gratefully acknowledges a senior postdoctoral fellowship from the Research Foundation Flanders (FWO) with grant agreement no. 1286521N. M.E.S. acknowledges the financial support provided by the Annie Jump Cannon Fellowship, supported by the University of Delaware and endowed by the Mount Cuba Astronomical Observatory.
This paper includes data collected with the TESS mission, obtained from the MAST data archive at the Space Telescope Science Institute (STScI). Funding for the TESS mission is provided by the NASA Explorer Program. STScI is operated by the Association of Universities for Research in Astronomy, Inc., under NASA contract NAS 5–26555. 
%Funding for the TESS Asteroseismic Science Operations Centre is provided by the
%Danish National Research Foundation (Grant agreement no.: DNRF106), ESA PRODEX
%(PEA 4000119301) and Stellar Astrophysics Centre (SAC) at Aarhus University.
We thank the \textit{TESS} and TASC/TASOC teams for their support of the
present work.
This research has made use of the SIMBAD database, operated at CDS, Strasbourg,
France.
Some of the data presented in this paper were obtained from the Mikulski
Archive for Space Telescopes (MAST).
STScI is operated by the Association of Universities for Research in Astronomy,
Inc., under NASA contract NAS5-2655.
Part of the analysed spectra has been obtained with the spectropolaimeter ESPaDOnS at the Canada-France-Hawaii Telescope (CFHT) which is operated by the National Research Council of Canada, the Institut National des Sciences de l'Univers of the Centre National de la Recherche Scientifique of France, and the University of Hawaii. Part of the used spectra has been aquired with the spectropolaimeter NARVAL at the T\'{e}lescope Bernard Lyot operated by the Observatoire Midi Pyr\'{e}n\'{e}es.
Some of the spectra have been obtained with the Mercator Telescope, operated on the island of La Palma by the Flemmish Community, at the Spanish Observatorio del Roque de los Muchachos of the Instituto de Astrof\'{i}sica de Canarias. The work of the HERMES spectrograph is supported by the Research Foundation - Flanders (FWO), Belgium, the Research Council of KU Leuven, Belgium, the Fonds National de la Recherche Scientifique (F.R.S.-FNRS), Belgium, the Royal Observatory of Belgium, the Observatoire de Gen\'{e}ve, Switzerland and the Th\"{u}ringer Landessternwarte Tautenburg, Germany.

\section*{Data Availability Statement}

The data underlying this article are available in the article and in its online supplementary material. The \textit{TESS} data were extracted from the MAST archive https://archive.stsci.edu. ESPaDOnS spectra used in this article are available at the CFHT archive maintained by the CADC https://www.cadc-ccda.hia-iha.nrc-cnrc.gc.ca/en/cfht.

\bibliographystyle{mnras}
\bibliography{Kobzar}

\appendix                                     
\section{Magnetic field measurements}
\begin{table*}
\begin{center}
\caption{Mean longitudinal magnetic field measurements for eight ApBp stars for this study.  }
\label{appendix}
\def\arraystretch{0.98}
\setlength{\tabcolsep}{3pt}
\begin{tabular}{lllclclc}
\hline
Name (HD) & Epoch of & BJD & $\langle B_{\rm z}\rangle$ G & BJD& $\langle B_{\rm z}\rangle$ G & BJD& $\langle B_{\rm z}\rangle$ G  \\
& zero phase & & & & & \\
\hline
10840 & 2458323.56575 & \multicolumn{2}{c}{\citet{Bagnulo2015}} & & & & \\
&&2453184.834907& -174$\pm$42 & &  \\

22920 & 2458409.05852 &\multicolumn{2}{c}{\citet{Borra1983}} & \multicolumn{2}{c}{ \citet{Shultz2022} (NARVAL)} & \multicolumn{2}{c}{ \citet{Shultz2022} (ESPaDOnS)}\\
&&2443736.889& 380$\pm$140 & 2458395.67899 & 487.13 $\pm$ 22.66 & 2456557.94247 & 230.78 $\pm$ 11.64 \\
&&2443737.877& 200$\pm$120 & 2458386.64070 & 148.42 $\pm$ 17.39 & 2456560.95704 & 481.11 $\pm$ 12.22 \\
&&2443738.847& 185$\pm$185 & 2458387.67402 & 463.73 $\pm$ 19.67 & 2456675.70610 & 451.58 $\pm$ 14.66 \\
&&2443740.864& 360$\pm$180 & 2458389.66201 & 188.29 $\pm$ 13.15 & 2457437.76397 & 383.64 $\pm$ 15.52 \\
&& \multicolumn{2}{c}{\citet{Mathys1997}} & & & & \\
&&2448847.874& 351$\pm$162 & & & &  \\

24712 & 2458441.95096 & \multicolumn{2}{c}{\citet{Ryabchikova1997}}  &\multicolumn{2}{c}{\citet{Leone2000}}  &  \multicolumn{2}{c}{\citet{Rusomarov2013}}\\
&& 2445534.924&  1440 $\pm$ 170  &2451115.558&  540$\pm$335 & 2455200.719667 & 742 $\pm$ 6\\
&& 2445535.919&  1620 $\pm$ 170  & 2451117.540&  530$\pm$125 & 2455201.735957 & 946 $\pm$ 7 \\
&& 2445537.907&  810 $\pm$ 240  & 2451118.528&  780$\pm$210 & 2455202.661397 & 1064 $\pm$ 7 \\
&& 2445540.929&  590 $\pm$ 190  & 2451119.522&  400$\pm$360 & 2455203.670667 & 1059 $\pm$ 6 \\
&& 2445541.912&  500 $\pm$ 140  & 2451120.514& -180$\pm$120 &  2455204.656307 & 912 $\pm$ 6 \\
&& 2445543.916&  690 $\pm$ 260  & 2451121.520& -225$\pm$220 &  2455205.681177 & 720  $\pm$6 \\
&& 2445544.908&  1000$\pm$ 150  & 2451215.301&  355$\pm$530 & 2455206.656887 & 512 $\pm$ 5\\
&& & &2451238.257&  550$\pm$460  & 2455209.652957 & 219 $\pm$ 3\\
&&\multicolumn{2}{c}{\citet{Mathys1994}} & \multicolumn{2}{c}{\citet{Wade2000}} & 2455210.670757 & 305 $\pm$ 4  \\
&& 2447189.549& 212$\pm$202  & 2450857.334&  610$\pm$ 43 & 2455211.669147 & 465 $\pm$ 5\\
&& 2447190.529& 709$\pm$236  &2451192.316& 1043$\pm$ 58 & 2455212.667537 &640 $\pm$ 6\\
&& \multicolumn{2}{c}{ This study (ESPaDOnS)} & 2451193.421&  653$\pm$ 86 & 2455213.671667 & 837 $\pm$ 6 \\ 
&&2456559.958 & 798 $\pm$ 20 & 2451197.362&  172$\pm$ 42 & 2455421.896510 & 251 $\pm$ 3 \\
&& 2456560.973 & 897 $\pm$ 25  & &   & 2455606.535706 & 321 $\pm$ 4 \\
&& & & & & 2455607.554866 & 223 $\pm$ 3\\

38170 & 2458438.05343 & \multicolumn{2}{c}{\citet{David-Uraz2021}} & & & & \\
&& 2458741.145913& -6$\pm$25 & &  & & \\
&& 2458743.146673&  105$\pm$14 & &  & & \\

63401 & 2458491.71207 & \multicolumn{2}{c}{\citet{Bagnulo2015}} & \multicolumn{2}{c}{This study (ESPaDOnS)}  & & \\
&& 2452678.529602& -589$\pm$53 &2455521.6568 & 308$\pm$65  & & \\
&& 2453002.556151&  153$\pm$95 &2457108.2348 & -109$\pm$58  & & \\
&& 2453004.731597& -414$\pm$101 &2457109.2598 & 396$\pm$60  & & \\
&& 2453399.628828&  322$\pm$55 & 2457110.2218 & 339$\pm$49  & & \\
&&               &             & 2457111.2948 & -47$\pm$23  & & \\
&&               &             & 2457112.2818 & 338$\pm$54  & & \\
&&               &             & 2457113.2938 & -400$\pm$70  & & \\
74521 & 2458491.78360 & \multicolumn{2}{c}{\citet{Bohlender1993}}& \multicolumn{2}{c}{\citet{Mathys1994}}  & \multicolumn{2}{c}{\citet{Leone2007} } \\
&& 2446834.679& 770$\pm$140& 2446894.528& 967$\pm$121 & 2453718.717& 1032$\pm$ 30 \\
&& 2446835.715& 640$\pm$140& 2446895.485& 850$\pm$128 & 2453719.710& 794$\pm$ 28 \\
&& 2446836.750& 580$\pm$140& 2447279.531& 398$\pm$190 & \multicolumn{2}{c}{This study  (ESPaDOnS)} \\
&& 2446837.703& 500$\pm$160& 2447280.529& 418$\pm$156 & 2457435.932 & 528$\pm$ 23 \\
&& 2446894.616& 200$\pm$300& 2447281.465& 603$\pm$107 & 2458437.606 & 541$\pm$ 25 \\
&& &                  & &      & 2458457.643 & 538$\pm$ 26  \\
77314 & 2458519.36909 & \multicolumn{2}{c}{This study (dimaPol)} & &  & & \\
&& 2455261.87680&-159$\pm$263& &  & & \\
&& 2455264.79807&259$\pm$220& &  & & \\
&& 2455580.93253&35$\pm$205& &  & & \\
&& 2456678.93416&-291$\pm$167 & &  & & \\
86592 & 2458573.33597 & \multicolumn{2}{c}{This study (dimaPol)} & \multicolumn{2}{c}{This study (dimaPol)}  & \multicolumn{2}{c}{This study (dimaPol)} \\
&& 2456355.84065&187$\pm$755& 2457088.84276&551$\pm$159  & 2458561.81803&664$\pm$447 \\
&& 2456992.07270&2323$\pm$297& 2457089.84700&1281$\pm$174  & 2458562.81450&2080$\pm$359 \\
&& 2456995.09549&2098$\pm$247& 2457090.87518&1579$\pm$143 & 2458563.81851&1190$\pm$449 \\
&& 2457084.84742&2421$\pm$181& 2457465.83889&2852$\pm$294 & 2458569.79994&873$\pm$238\\
&& 2457085.87101&505$\pm$248& 2457476.78068&1613$\pm$121  & & \\
\hline

\end{tabular}
\end{center}
\end{table*}

\bsp	% typesetting comment
\label{lastpage}
\end{document}